\newcommand{\simnot}{\mathord{\sim}}
\documentclass[journal]{IEEEtran}
\IEEEoverridecommandlockouts
\usepackage{cite}
\usepackage{comment}
\usepackage{amsmath,amssymb,amsfonts}

\usepackage{algpseudocode}
\usepackage{graphicx}
\usepackage{subfigure}
\usepackage{float}
\usepackage{subfig}
\usepackage[font=small,labelfont=rm]{caption}
\usepackage{textcomp}
\usepackage{xcolor}
\usepackage[linesnumbered,ruled,vlined]{algorithm2e}
\SetKwInput{KwInput}{Input}                
\SetKwInput{KwOutput}{Output}              

\def\BibTeX{{\rm B\kern-.05em{\sc i\kern-.025em b}\kern-.08em
    T\kern-.1667em\lower.7ex\hbox{E}\kern-.125emX}}

\begin{document}

\title{Reliability and Delay Analysis of 3-Dimensional Networks with Multi-Connectivity: Satellite, HAPs, and Cellular Communications}

\author{\IEEEauthorblockN{Fateme Salehi\IEEEauthorrefmark{1}\IEEEauthorrefmark{2}, Mustafa Ozger\IEEEauthorrefmark{1}, Cicek Cavdar\IEEEauthorrefmark{1}}\\
\IEEEauthorblockA{\IEEEauthorrefmark{1}School of Electrical Engineering and Computer Science,  KTH Royal Institute of Technology, Sweden \\\IEEEauthorrefmark{2}Department of Computer and Electrical Engineering, Mid Sweden University, Sweden}
}

\maketitle

\begin{abstract}
Aerial vehicles (AVs) such as electric vertical take-off and landing (eVTOL) aircraft make aerial passenger transportation a reality in urban environments. However, their communication connectivity is still under research  to realize their safe and full-scale operation. This paper envisages a multi-connectivity (MC) enabled aerial network to provide ubiquitous and reliable service to AVs. Vertical heterogeneous networks with  direct air-to-ground (DA2G) and air-to-air (A2A) communication, high altitude platforms (HAPs), and low Earth orbit (LEO) satellites are considered. We evaluate the end-to-end (E2E) multi-hop reliability and network availability of the downlink of AVs for remote piloting scenarios, and control/telemetry traffic. Command and control (C2) connectivity service requires ultra-reliable and low-latency communication (URLLC), therefore we analyse  E2E reliability and latency under the finite blocklength (FBL) regime. We explore how different MC options satisfy the demanding E2E connectivity requirements taking into account antenna radiation patterns and unreliable backhaul links. Since providing seamless connectivity to AVs is very challenging due to the line-of-sight (LoS) interference and reduced gains of downtilt ground base station (BS) antennas, we use coordinated multi-point (CoMP) among ground BSs to alleviate the inter-cell interference. 
Furthermore, we solve an optimization problem to select the best MC path under the quality of service (QoS) constraints. We maximize spectral efficiency (SE) to specify the optimum MC path with the minimum number of required links.
Based on the simulation results, we find out that even with very efficient interference mitigation, MC is the key enabler for safe remote piloting operations.
\end{abstract}

\begin{IEEEkeywords}
reliability, network availability, multi-connectivity, aerial vehicles, URLLC,  coordinated multi-point.
\end{IEEEkeywords}

\section{Introduction}\label{sec.intro}
\IEEEPARstart
{F}{uture} aerial communications (FACOM) is defined as the connectivity ecosystem incorporating emerging aerial use cases with different aerial vehicles (AVs) and range of connectivity solutions\cite{FACOM} such as high altitude platforms (HAPs), air-to-air (A2A), direct air-to-ground (DA2G) communication, and satellites. One critical emerging scenario is remote piloting of AVs where connectivity has a key role to ensure the safe operations. In command and control (C2) communication links, short-packet control information needs to be transmitted with ultra-reliable and low-latency communications (URLLC). AVs have different mission and flight characteristics with diverse quality of service (QoS) requirements such as data rate, end-to-end (E2E) latency and communication reliability. AVs such as flying taxis and electric vertical take-off and landing (eVTOL) enable passengers to be transported over several tens of kilometers at low altitudes as an extension of the urban transportation system \cite{FACOM}. 

There is a growing interest in the design and performance analysis of FACOM to provide connectivity to AVs through different technologies. Performance of DA2G communication is studied to connect AVs directly with the ground cellular networks through beamforming and 5G networks \cite{DA2GCMag, DA2GCDeploy}. Network architectures and business models to provide high capacity DA2G communication is studied for passenger aircraft use case \cite{DA2GC-business
}. A scenario for beyond visual line-of-sight (BVLoS) operation for remote piloting of unmanned aerial vehicles (UAVs) in sub-6 GHz \cite{BVLoS} and millimeter waves \cite{BVLoS_mmWave} is studied, which utilizes different technologies such as mobile edge computing and augmented reality. Multi-hop A2A communication is considered in \cite{aircraft} to extend DA2G communication without considering the reliability performance. In \cite{hybrid_TerrestAerial}, the authors propose macro-diversity scheme considering a terrestrial and aerial hybrid network for ensuring URLLC services under the centralized RAN with ideal backhaul.
The authors of \cite{URLLC_UAV} exploit the macro-diversity gain of the distributed multi-antenna systems and the array gain of the centralized multi-antenna systems. They maximize the availability of the C2 communication links between UAVs and a ground base station (BS) by optimizing the altitude of UAVs, the duration of the uplink and downlink phases, and the antenna configuration. 

HAPs communication is envisioned to be a part of non-terrestrial networks (NTNs) to ensure continuous, ubiquitous and scalable services \cite{haps_ntn}. HAPs may be used in several use cases ranging from extension of terrestrial coverage for white spot areas to disaster recovery support. They offer potential benefits such as high capacity links with large footprints and favorable line of sight (LoS) link conditions and computation offloading not only in suburban but also urban areas \cite{Kurt21vision}. Beyond DA2G and A2A communications, HAPs communication serves as a connectivity option for FACOM.  In addition to providing connectivity to terrestrial users, HAPs enable reliable connectivity to AVs \cite{HAPS_SMBS}. One use case of HAPs is to support highly reliable and low latency communication for remote piloting of AVs in a multi-link connectivity setting \cite{MC-AV}. Furthermore, HAPs have more computational power than AVs, hence they can provide an intelligence layer in the sky for AVs \cite{Kurt21vision}.

One of the methodologies to provide URLLC services without intervention in the physical layer design is to utilize multi-connectivity (MC). MC by introducing link/path diversity can improve both latency and reliability performance. There are various architectures for MC with different means of diversity such as BS diversity, network diversity, and technology diversity. Coordinated multi-point (CoMP) architecture belongs to the first category, i.e., BS diversity, where multiple BSs from the same network simultaneously serve an AV to improve the overall communication reliability. In this regard, the authors of \cite{3D_CoMP} propose a 3D CoMP model for A2A communication, where UAVs were employed both as aerial BSs as well as aerial UEs. The authors of \cite{CoMP_UAV} use CoMP transmission for providing seamless connectivity to UAVs, and the coverage probability is studied for two scenarios with static hovering UAVs and mobile UAVs. In \cite{CoMP_sky}, CoMP in the sky is proposed for uplink communications and UAV placement and movement are optimized to maximize the network throughput. None of the above works consider CoMP for URLLC with E2E performance analysis.

For MC with network diversity we can refer \cite{drone_dc} and \cite{dual_MNO}, where the authors conduct field measurements with a UAV to evaluate the improvements in reliability and latency over multiple mobile network operators (MNOs). They also report performance gain with multiple links compared with single link due to the performance variations of the MNOs at different altitudes and environments.
Moreover, in \cite{maritime}, a combination of a public and dedicated cellular network with multipath transmission control protocol (MPTCP) is proposed for maritime search and rescue missions of UAVs. The results show that the multi-link protocol increases the range and improves the data rate performance.

MC can also be considered using different radio access technologies (RATs). The authors of \cite{multipath} aim to provide robust bandwidth allocation for retaining the continuous and stable connectivity among dynamic system components. To this end, they present an analytic modeling of MPTCP with a satellite link and WiFi access points to control a swarm of UAVs without guaranteeing the reliability and delay requirements. The authors of \cite{3fold_redundant} present field measurements of triple-redundant multi-link architecture employing cellular, WiFi and LoRa for the C2 link connectivity with communication range and latency performance criteria. Their redundancy design employs a cellular network as the primary link and the other two as fallback links when there is no cellular coverage. None of the previous studies consider E2E paths for C2 communications containing backhaul links and network architectures.
In \cite{MC-AV}, we consider a heterogeneous network of ground BSs, relay AVs, and a HAP to provide connectivity for AVs. We take into account practical antenna configurations with unreliable backhaul links. The automatic repeat-request (ARQ) mechanism and frequency diversity is employed to improve reliability of radio links. Mean-value analysis of E2E reliability and latency is considered for the performance evaluation.

Reliability is a critical metric in BVLoS control of AVs. In this paper we capture both error rate and delay analysis, while we also define a service specific availability metric. Network availability is defined as the probability that both reliability and delay requirements can be met simultaneously. 
In our study, different from prior works in \cite{DA2GCMag,DA2GCDeploy,DA2GC-business,BVLoS,BVLoS_mmWave,aircraft,hybrid_TerrestAerial,URLLC_UAV,3D_CoMP,CoMP_UAV,CoMP_sky,drone_dc,dual_MNO,maritime,multipath,3fold_redundant}, we aim to investigate the minimum required connectivity links and spectrum for the safe and full-scale remote piloting operation of BVLoS.
As concepts of eVTOLs are of recent venture, to the best of our knowledge, the literature has not yet covered the connectivity needs and potential solutions for the C2 communication links of these aerial platforms. In this regard, we consider a rigorous analysis of E2E delay and reliability of communication paths, which includes different delay and error parameters in wired backhaul links, transmitter's queue, and wireless links with small- and large-scale fading. 3-Dimensional (3D) MC consisting of DA2G, A2A, HAP, and low Earth orbit (LEO) satellite communications is considered as the enabler of stringent requirements of remote piloting operation. Additionally, to improve the reliability of DA2G communication we exploit CoMP in joint transmission (JT) mode among ground BSs. We characterize the effect of different parameters such as data rate, bandwidth, CoMP cluster size, interference, and backhaul failure on the latency, reliability, and network availability and finally investigate how multi-path connectivity of RAT diversity can guarantee the requirements for safe operation. 

The main contributions of this paper can be summarized as follows.
\begin{itemize}
    \item We consider RAT diversity of DA2G, A2A, HAP, and LEO satellite to provide seamless connectivity for remote piloting of AVs.
    \item We utilize ground BS diversity, namely JT CoMP, for interference mitigation of DA2G communication and increase reliability.
    \item We present the E2E analysis of latency and reliability of C2 communications under finite blocklength (FBL) regime, and automatic repeat-request (ARQ) mechanism is employed to improve reliability of radio links. 
    \item We consider network architecture with unreliable backhaul links and buffer queues, as well as, practical antenna configurations for ground BSs with downtilt antennas and HAP/satellite with multi-beam antenna patterns.
    \item We compare the performance of different MC options from reliability and network availability perspective through numerical analysis with extensive Monte-Carlo simulations.
    \item We solve an optimization problem based on the brute-force method to find the best MC path with a minimum number of links enable to ensure the QoS requirements.
\end{itemize}

The paper is organized as follows. Section \ref{sec.sys} presents the system model consisting of the considered scenario, key performance indicators (KPIs), and the methodology. Section \ref{sec.chmodel} presents the channel models of communication links, antenna radiation pattern of ground BSs and HAP/satellite, and SINR calculation. Section \ref{sec.anal} presents E2E latency and reliability analysis of different RATs and communication paths. Section \ref{sec.res} discusses the numerical results and investigate the requirements with the MC options to enable remote piloting operation in different system parameters. Section \ref{sec.con} concludes the paper.

\section{System Model}\label{sec.sys}
In this section, we introduce the considered scenario with its requirements and related KPIs as well as MC as the methodology for providing them. 
\subsection{Remote Piloting of AVs and QoS Requirements}
BVLoS remote piloting of an AV requires a communication path between the remote pilot and the AV. In this concept, ground pilots remotely navigate an AV, which can supply pilots with a first-person view by on-board cameras and other useful sensor data. Remote piloting operation emphasizes the demand for resilient E2E communication paths from the remote pilots to the AVs. As eVTOLs and UAVs occupy the sky, they must coordinate with one another as well as other AVs to efficiently share the low-altitude sky. Unmanned traffic management (UTM) introduce the regulation of these vehicles in a more-autonomous manner compared with the air traffic management (ATM). Machine-type communications (MTC) can become the dominant connectivity type in UTM rather than the human-centric ATM communication in the future \cite{FACOM}. 
Based on \cite{FACOM}, control/telemetry traffic for remote piloting operations of eVTOLs requires a data rate about $0.25\sim1$ Mbps, E2E latency less than $10\sim150$ ms, and the minimum communication reliability $99.999\%$.

\subsection{Key Performance Indicators}\label{sub.sysKPI}
The most important KPIs related to URLLC are latency, reliability, and network availability. \textbf{Latency} is defined as the delay a packet experiences from the ingress of a protocol layer at the transmitter to the egress of the same layer at the receiver \cite{popovski_URLLC}. In the URLLC literature, the \textbf{reliability} is reflected either by packet loss probability or by latency, which we call them error-based and delay-based reliability, respectively. The E2E packet loss probability, $\mathcal{E}_{\rm{E2E}}$, includes different components such as backhaul failure probability, queueing delay violation, decoding error probability, and so on. Therefore, in \emph{error-based reliability}, the reliability requirement which is defined by
\begin{equation}
\label{eq.errorReliabil}
    \mathcal{R}=1-\mathcal{E}_{\rm{E2E}}\ , 
\end{equation}
can be satisfied if the overall packet loss probability does not exceed $\varepsilon^{\rm{th}}$. On the other hand, using the convention that dropped packets have infinite latency, authors of \cite{popovski_URLLC} define the reliability as the probability that the latency does not exceed a pre-defined threshold $D^{\mathrm{th}}$. Thus, in \emph{delay-based reliability}
\begin{equation}
\label{eq.delayReliabil}
    \mathcal{R}=\operatorname{Pr}\left \{\mathcal{D}_{\rm{E2E}} \leq D^{\mathrm{th}} \right \},
\end{equation}
where $\mathcal{D}_{\rm{E2E}}$ is the E2E delay from the transmitter to the receiver. 

Different from latency and reliability, which are the QoS required by each user, \textbf{availability} captures the performance of the network how it can respond to the demands of the users, and is another key performance metric for URLLC. In the conventional systems, availability is specified by the packet loss probability which we call it \emph{error-based network availability}, i.e.,
\begin{equation}
\label{eq.errorAvailabil}
    P_{\rm{A}}=\operatorname{Pr}\left \{\mathcal{E}_{\rm{E2E}} \leq \varepsilon^{\rm{th}} \right \}.
\end{equation}
However, for URLLC services, availability is defined as the probability that the network can support a service with a target QoS requirement on both latency and reliability \cite{she_magazine}. Based on the above definitions, the availability for URLLC services can be described by the following equation which we call it as \emph{delay-aware network availability}
\begin{equation}
\label{eq.availability}
    P_{\rm{A}}=\operatorname{Pr}\left \{\mathcal{E}_{\rm{E2E}} \leq \varepsilon^{\rm{th}},\mathcal{D}_{\rm{E2E}} \leq D^{\mathrm{th}} \right \}.
\end{equation}
Here $\varepsilon^{\rm{th}}$ and $D^{\mathrm{th}}$ characterize the QoS requirements in terms of packet error and delay.

\subsection{Multi-Connectivity}
\begin{figure}[b]
    \centering
            \includegraphics[width=1\columnwidth]{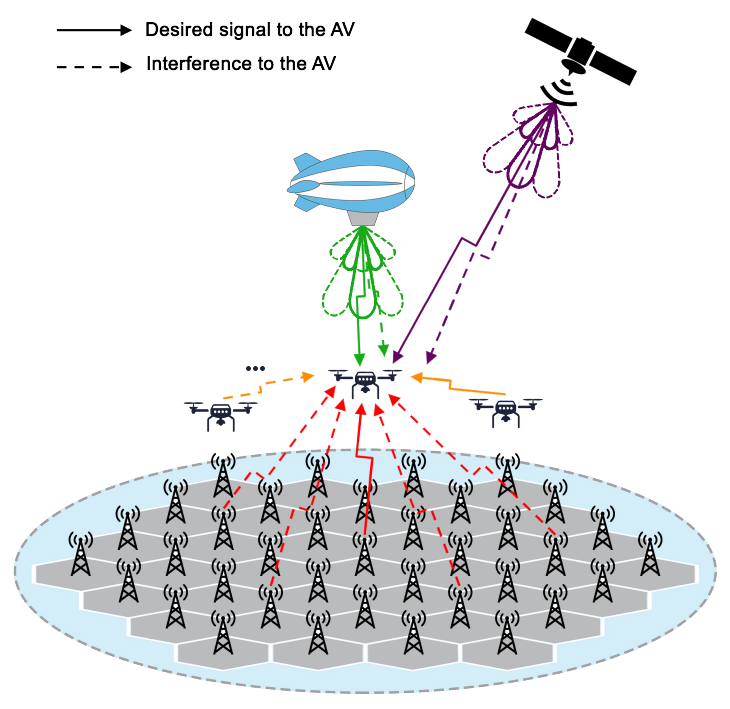}
            \caption{System model.} 
            \label{fig.sys}
\end{figure}

\begin{figure}[t]
    \centering
            \includegraphics[width=1\columnwidth]{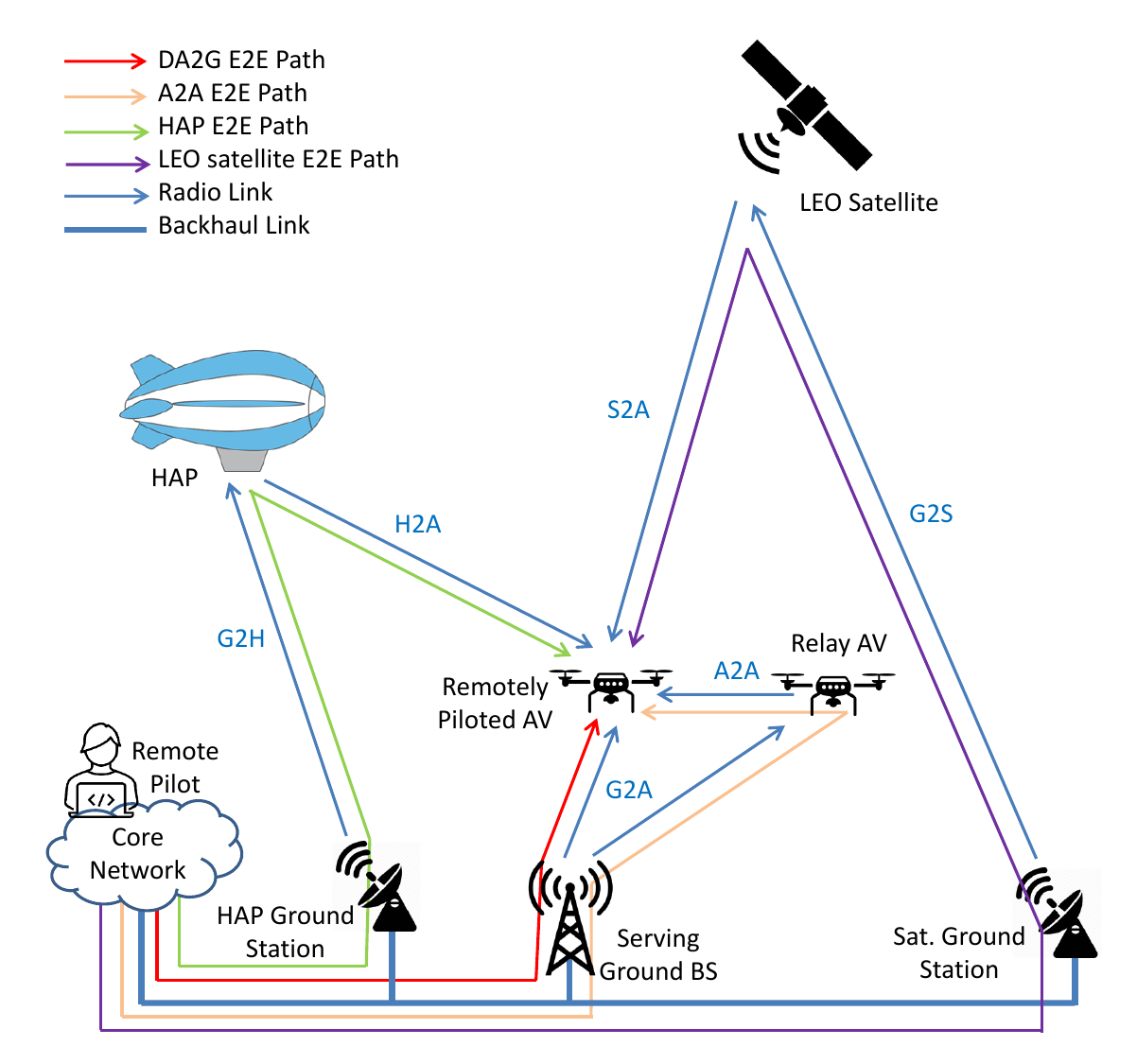}
            \caption{Illustration of multi-RAT and E2E communication paths.} 
            \label{fig.path}
\end{figure}

MC using multiple communication paths simultaneously is the key technology to reduce latency and increase reliability to fulfill strict requirements of AVs' remote piloting. As shown in Fig. \ref{fig.sys}, the system model consists of an integration of multiple RATs including DA2G, A2A, HAP, and LEO satellite communication. For all the RATs, we assume particular frequency band with full frequency reuse such that each link experiences probabilistic interference from all the corresponding links. The E2E path of each RAT is illustrated in Fig. \ref{fig.path}, a directive path starting with the core network, traversing the backhaul link and the radio link (downlink) to reach the destination AV, which is the AV that remote pilot wants to navigate. The communication links consist of ground BS-to-AV (G2A), HAP ground station-to-HAP (G2H), satellite ground station-to-LEO satellite (G2S), and AV/HAP/LEO satellite-to-AV (A2A/H2A/S2A). In Fig. \ref{fig.path}, four different E2E paths are shown, i.e., the red line which illustrates ``DA2G E2E path" includes the backhaul link to the ground BS and G2A link. ``A2A E2E path", illustrated with orange line is defined as the path consisting of backhaul, G2A and A2A links. The green line illustrates the ``HAP E2E path" defined as the path consisting of backhaul link to the HAP ground station, G2H and H2A links. Finally, the ``LEO satellite E2E path" indicated with violet line includes the backhaul link to the satellite ground station, G2S and S2A links.

\subsection{Transmission and Combining Strategy}
We consider packet cloning for transmitting the message from the remote pilot to the AV over independent links. In this approach, the source sends copies of the message through each of the available links \cite{interface_diversity}. The combining scheme is joint decoding, where each link is decoded individually. Thus, the overall packet loss probability of $N$ parallel transmission paths is 
\begin{equation}
\label{eq.error.tot}
    \mathcal{E}_{\rm{E2E}}=\prod_{i=1}^{N} \mathcal{E}_{\rm{E2E}}^{i},
\end{equation}
where $\mathcal{E}_{\rm{E2E}}^{i}$ is the error probability of the $i$th path, and $i\in \{\rm{g},\rm{a},\rm{h},\rm{s}\}$ refers to different RATs including DA2G, A2A, HAP, and satellite communications, respectively. It also potentially reduces the delay, since only the packet that arrives earlier and is decoded correctly needs to be considered. Hence, the E2E delay of multi-RAT using the cloning scheme is calculated as \cite{interface_diversity}
\begin{equation}
\label{eq.delay.tot}
    \mathcal{D}_{\rm{E2E}}=\min _{i=1,\dots,N} \left\{\mathcal{D}_{\rm{E2E}}^{i}\right\},
\end{equation}
where $\mathcal{D}_{\rm{E2E}}^{i}$ is the E2E delay of the $i$th path.

\section{Channel Models of Communication Links and Antennas Radiation Patterns}\label{sec.chmodel}
To model a realistic propagation channel, we consider both large-scale fading and small-scale fading.
\subsection{Large-Scale Fading}
\subsubsection{Path Loss of G2A Link}
We consider that the G2A link experiences LoS propagation with a probability of $P_\mathrm{LoS}$, which is calculated as \cite{U2U_comm}
\begin{equation}
  P_\mathrm{LoS}=\prod_{j=0}^{k}\left[1-\exp \left(-\frac{\left[\hslash_{\mathrm{g}}-\frac{(j+0.5)\left(\hslash_{\mathrm{g}}-\hslash_{\mathrm{a}}\right)}{k+1}\right]^{2}}{2 \mathrm{q}_{3}^{2}}\right)\right],  
\end{equation}
where $k=\left\lfloor\frac{r_\mathrm{ga} \sqrt{\mathrm{q}_{1} \mathrm{q}_{2}}}{1000}-1\right\rfloor$, and $r_\mathrm{ga}$ is the 2D distance between the ground BS and the AV, while $\{\mathrm{q}_{1}, \mathrm{q}_{2}, \mathrm{q}_{3}\}$ are environment-dependent parameters set to $\{0.3, 500, 20\}$ to model an urban scenario \cite{U2U_comm}. Moreover, $\hslash_{\mathrm{g}}$ and $\hslash_{\mathrm{a}}$ are the height of the ground BS and altitude of the AV, respectively.

Thus, the average path loss of the G2A link is derived as
\begin{equation}
\label{eq.PL_ga}
  \mathit{PL}_\mathrm{ga}=P_\mathrm{LoS} \times \mathit{PL}^{\mathrm{LoS}}_\mathrm{ga}+(1-P_\mathrm{LoS}) \times \mathit{PL}^{\mathrm{NLoS}}_\mathrm{ga},
\end{equation}
where $\mathit{PL}^{\mathrm{LoS}}_\mathrm{ga}$ and $\mathit{PL}^{\mathrm{NLoS}}_\mathrm{ga}$ are the path losses of the G2A channel under LoS and NLoS conditions, respectively. Based on the urban macro cells (UMa) scenario, $\mathit{PL}^{\mathrm{LoS}}_\mathrm{ga}$ and $\mathit{PL}^{\mathrm{NLoS}}_\mathrm{ga}$ are calculated as follows \cite{3GPP36.777}
\begin{equation}
    \mathit{PL}^{\mathrm{LoS}}_\mathrm{ga}(\mathrm{dB})=28+22\log_{10} \left(d_\mathrm{ga}\right) + 20\log_{10} \left( f_{\rm{c}} \right),
\end{equation}
\begin{equation}
\begin{split}
    \mathit{PL}^{\mathrm{NLoS}}_\mathrm{ga}(\mathrm{dB})&=-17.5+\left(46 - 7\log_{10} \left( \hslash_{\rm{a}} \right)\right)\log_{10} \left(d_\mathrm{ga}\right)\\
    &+ 20\log_{10} \left( 40\pi f_{\rm{c}}/3 \right),    
\end{split}
\end{equation}
where $d_\mathrm{ga}$ is the 3D distance between the ground BS and the AV in meter, and $f_{\rm{c}}$ is the carrier frequency in GHz.

\subsubsection{Path Loss of A2A/H2A/S2A Link}
For A2A, H2A, and S2A links, the free space path loss (FSPL) channel model is used \cite{3GPP38.811}
\begin{equation}
\begin{split}
     \mathit{PL}_{\rm{xy}}{(\rm{dB})} &= \mathit{FSPL}(d_{\rm{xy}},f_{\rm{c}})\\ &= 32.45 + 20 \log_{10} \left( d_{\rm{xy}} \right) + 20\log_{10} \left( f_{\rm{c}} \right),    
\end{split}
\end{equation}
where $\rm{xy}\in\{\rm{aa},\rm{ha},\rm{sa}\}$ represents the A2A, H2A, and S2A link, respectively. $d_{\rm{xy}}$ is the 3D distance between nodes $\mathrm{x}$ and $\mathrm{y}$ in meter, and $f_{\rm{c}}$ is the carrier frequency in GHz.

\subsubsection{Path Loss of G2H/G2S Links}
The path loss of the G2H/G2S link can be considered as the basic path loss model which accounts for the signal's FSPL, shadow fading (SF), and clutter loss (CL) \cite{3GPP38.811} 
\begin{equation}
    \mathit{PL}_{\rm{xy}}{(\rm{dB})} = \mathit{FSPL}(d_{\rm{xy}},f_{\rm{c}}) + \mathit{SF} + \mathit{CL}\ ,
\end{equation}
where $\rm{xy}\in\{\rm{gh},\rm{gs}\}$ represent the G2H link and the G2S link, respectively. SF is modeled by a log-normal distribution, i.e., $\mathit{SF}\sim N(0,\sigma_{\mathit{SF}}^2)$. CL based on \cite[Table 6.6.2-1]{3GPP38.811} depends on the elevation angle between nodes $\mathrm{x}$ and $\mathrm{y}$, the carrier frequency, and the environment. When there is LoS condition, CL is negligible and can be considered as 0 dB in the basic path loss model \cite{3GPP38.811}.

\subsection{Small-Scale Fading}
Due to the LoS path for all the mentioned links, small-scale channel fading between nodes $\mathrm{x}$ and $\mathrm{y}$, i.e., $\omega_\mathrm{xy}$, can be taken into account as the Rician model, where $\rm{xy} \in \{\rm{ga},\rm{aa},\rm{gh},\rm{ha},\rm{gs},\rm{sa}\}$.
\begin{equation}
f_{\Omega}(\omega_\mathrm{xy})=\frac{\omega_\mathrm{xy}}{\sigma_\mathrm{xy}^{2}} \exp \left(\frac{-\omega_\mathrm{xy}^{2}-\rho_\mathrm{xy}^{2}}{2 \sigma_\mathrm{xy}^{2}}\right) I_{0}\left(\frac{\omega_\mathrm{xy} \rho_\mathrm{xy}}{\sigma_\mathrm{xy}^{2}}\right),
\end{equation}
with $\omega_\mathrm{xy} \geq 0$, and $\rho_\mathrm{xy}$ and $\sigma_\mathrm{xy}$ reflecting the strength of the LoS and the NLoS paths, respectively. $I_{0}(.)$ denotes the modified Bessel function of the first kind and zero order. The Rice factor of X2Y link, $K_\mathrm{xy}$, is defined as
 \begin{equation}
     K_\mathrm{xy}({\rm{dB}})=10\log_{10}\left(\frac{\rho_\mathrm{xy}^{2}}{2 \sigma_\mathrm{xy}^{2}}\right),
 \end{equation}
which increases directly with different parameters such as altitude, elevation angle, and carrier frequency. The elevation angle plays a dominant role among the other factors \cite{Rice_factor}.

\subsection{Antenna Gain}
We assume that all AVs are equipped with a single omnidirectional antenna with unitary gain. However, we consider realistic antenna radiation patterns for the ground BSs and the HAP/satellite, which are given as follows. 
\subsubsection{Ground BS Antenna Pattern}
We assume that the ground BSs are equipped with a vertical, $N_{\rm{e}}$-element uniform linear array (ULA), where each element is omnidirectional in azimuth with a maximum gain of $g_{\rm{e}}^{\max}$ and directivity as a function of the zenith angle $\phi$ \cite{U2U_comm}:
\begin{equation}
g_{\rm{e}}(\phi)=g_{\rm{e}}^{\max } \sin ^{2} \phi.
\end{equation}
We assume that there is a half-wavelength spacing between the adjacent antenna elements. With a fixed downtilt angle $\phi_{\rm{t}}$, the array factor of the ULA is given by \cite{U2U_comm}
\begin{equation}
g_{\rm{A}}(\phi)=\frac{\sin ^{2}\left(N_{\rm{e}} \pi\left(\cos \phi-\cos \phi_{\mathrm{t}}\right) / 2\right)}{N_{\rm{e}} \sin ^{2}\left(\pi\left(\cos \phi-\cos \phi_{\mathrm{t}}\right) / 2\right)}.
\end{equation}
The total ground BS's antenna gain in linear scale is 
\begin{equation}
  g_{\rm{g}}(\phi)=g_{\rm{e}}(\phi)\times g_{\rm{A}}(\phi).
\end{equation}

\subsubsection{HAP/Satellite Antenna Pattern}
For HAP and satellite, multi-beam antennas instead of uniform planar array antennas are considered, as uniform planar array configuration requires the design of a precoding matrix which is beyond the scope of this paper. It is assumed that each cell is served by one main beam \cite{HAP,Sat}. The following normalized antenna $\rm{x}$ gain pattern of one beam, $\rm{x} \in \{\rm{h},\rm{s}\}$, corresponding to a typical reflector antenna with a circular aperture with a radius of 10 wavelengths, is considered \cite{3GPP38.811}
\begin{equation}
g_{\rm{x}}(\theta)=\left\{\begin{array}{ll}
1, & \text { for } \theta=0, \\
4\left|\frac{J_{1}(20\pi \sin \theta)}{20\pi \sin \theta}\right|^{2}, & \text { for } 0<|\theta| \leq 90^{\circ},
\end{array}\right.
\end{equation}
where $\theta$ is the angle with respect to antenna boresight, and $J_1(.)$ is the Bessel function of the first kind and first order.

\subsection{SINR Calculation}
One may obtain the channel coefficient between any two nodes $\mathrm{x}$ and $\mathrm{y}$ as
\begin{equation}
    h_\mathrm{xy} = (\frac{g_{\rm{xy}}}{\mathit{PL}_\mathrm{xy}})^{1/2}\omega_\mathrm{xy},
\end{equation}
where $g_{\rm{xy}}$ is the total antenna gain between nodes $\mathrm{x}$ and $\mathrm{y}$ given by the product of their respective antenna gains. Finally, the SINR of X2Y link with bandwidth $B^{\rm{xy}}$, $\rm{xy} \in \{\rm{ga},\rm{aa}\}$, is calculated as follows
\begin{equation}
    \gamma^{\rm{xy}} = \frac{p_{\rm{x}}|h_\mathrm{xy}|^2}{P_{\rm{interf}}\sum\limits_{i \in \mathcal{N}_i} {p_{\mathrm{x}_i}} |{h_{\mathrm{x}_i\mathrm{y}}}{|^2}+B^{\rm{xy}}N_0},
\end{equation}
where $p_{\rm{x}}$ is the transmit power of node $\mathrm{x}$, and $N_0$ is the noise spectral density. $\mathcal{N}_i$ is the set of interfering nodes and, $h_{\mathrm{x}_i\mathrm{y}}$ indicates the channel coefficient between the interfering node $\mathrm{x}_i$ and node $\mathrm{y}$. 
We assume that interference cancellation techniques can harness interference \cite{deepSIC,selfIC,d2dIC}, and it can be explicitly captured by interference probability denoted by $P_{\rm{interf}}$. It points out that the higher the interference cancellation, the lower the interference probability. Hence, the effect of interference power on the network is affected by $P_{\rm{interf}}$ due to the fact that each potential interferer is modeled as a Bernoulli random variable with a probability of $P_{\rm{interf}}$.
We also assume that the G2H and the G2S links are interference-free, while the interference on H2A/S2A links is due to the side lobes of HAP/satellite's antenna overlapping with the main lobes \cite{HAP,Sat}.

\section{Reliability and Latency Analysis}\label{sec.anal}
\subsection{Preliminaries}
\subsubsection{Transmission Analysis in the FBL Regime}
The achievable data rate of the X2Y link, $R^{\rm{xy}}$, with FBL coding and an acceptable Block Error Rate (BLER) $\varepsilon_{\rm{t}}^{\rm{xy}}$, $\rm{xy} \in \{\rm{ga},\rm{aa},\rm{gh},\rm{ha},\rm{gs},\rm{sa}\}$, has an approximation as \cite{jointUlDl}
\begin{equation}
    \label{eq.rate}
    R^{\rm{xy}}\approx B^{\rm{xy}}\left(C^{\rm{xy}}-\sqrt{\frac{V^{\rm{xy}}}{B^{\rm{xy}}D_{\rm{t}}^{\rm{xy}}}}\frac{Q^{-1}(\varepsilon_{\rm{t}}^{\rm{xy}})}{\ln2}\right)\text{bits/s}\ ,
\end{equation}
where $C^{\rm{xy}}=\log_2(1+\gamma^{\rm{xy}})$ is the Shannon capacity and $V^{\rm{xy}}=1-(1+\gamma^{\rm{xy}})^{-2}$  is the channel dispersion. Moreover, $D_{\rm{t}}^{\rm{xy}}$ is the transmission delay of the X2Y link, and $Q^{-1}(\cdot)$ refers to the inverse Gaussian Q-function $Q(x)=\frac{1}{\sqrt{2\pi}}\int_{x}^{\infty} e^{-\frac{t^2}{2}} \,\mathrm{d}t$.

In the FBL regime, decoding error probability is given by
\begin{equation}
    \label{eq.tr.error}
    \varepsilon_{\rm{t}}^{\rm{xy}}\approx Q\left(f(\gamma^{\rm{xy}},R^{\rm{xy}},D_{\rm{t}}^{\rm{xy}})\right)\ ,
\end{equation}
where 
\begin{equation}
f(\gamma^{\rm{xy}},R^{\rm{xy}},D_{\rm{t}}^{\rm{xy}})\triangleq\frac{\left(B^{\rm{xy}}C^{\rm{xy}}-R^{\rm{xy}}\right)\ln2}{\sqrt{B^{\rm{xy}}V^{\rm{xy}}/D_{\rm{t}}^{\rm{xy}}}}\ .
\end{equation}
When transmitting a packet that contains $b$ bits over the allocated channel, the decoding error probability can be obtained by substituting $D_{\rm{t}}^{\rm{xy}}=\frac{b}{R^{\rm{xy}}}$ into \eqref{eq.tr.error}. The above expressions are for AWGN channels which contain no fading. Here, we can assume our channel as a quasi-static flat fading channel such that at each realization, its characteristics remain the same.
 
By adopting ARQ scheme, the packet is retransmitted until it is received correctly, and we assume that there is a reliable feedback from the AV to the transmitter as in \cite{ARQ}. Hence, the average transmission delay of the X2Y link is calculated as
\begin{equation}
    {\overline{D}_{\rm{t}}^{\rm{xy}}}=\frac{{D_{\rm{t}}^{\rm{xy}}}}{1-{\varepsilon}_{\rm{t}}^{\rm{xy}}}\ .
\end{equation}

\subsubsection{Queueing Analysis}
As stated in \cite{jointUlDl}, the packet arrival process to the BS in MTC, which is an aggregation of packets generated by multiple sensors, can be modeled as a Poisson process. The event that each sensor at any given instant has a packet to upload or not is modeled as a Bernoulli process. The probability that sensor $m$ has a packet to upload is denoted by $P_m$. Then, the arrival process to the BS is defined as a Poisson process, because the sensors are independent. Since MTC is the connectivity type in our scenario, each remote pilot resembles a sensor that at any time instant may deliver a packet to the AV of interest via node $\rm{x}$. Therefore, if assume that $M_{\rm{x}}$ AVs are served by node $\rm{x}$, where $\rm{x} \in \{\rm{g},\rm{a},\rm{h},\rm{s}\}$ refers to ground BS, relay AV, HAP, and LEO satellite, respectively, the average total arrival rate to node $\rm{x}$ is $\lambda_{\rm{x}}=\sum_{m=1}^{M_{\rm{x}}}P_m$ packets/s.

Denote the packet dropping probability due to queueing delay violation as
\begin{equation}
     \varepsilon_{\rm{q}}^{\rm{x}}=\operatorname{Pr}\left \{D_{\rm{q}}^{\rm{x}}>D_{\rm{q},\text{max}} \right \},
\end{equation}
where $D_{\rm{q}}^{\rm{x}}$ is the queue delay of node $\rm{x}$, and $\rm{x} \in \{\rm{g},\rm{a},\rm{h},\rm{s}\}$. 
As described above, the packet arrival process to node $\rm{x}$ can be modeled as a Poisson process with the average arrival rate of $\lambda_{\rm{x}}$ packets/s. Then, the effective bandwidth of node $\rm{x}$, which is the minimal constant packet service rate required to satisfy the queueing delay requirement $(D_{\rm{q},\text{max}},\varepsilon_{\rm{q}}^{\rm{x}})$ can be expressed as follows \cite{jointUlDl}
\begin{equation}
\label{eq.effBW}
E_{\mathrm{BW}}^{\rm{x}}=\frac{ \ln \left(1 / \varepsilon_{\mathrm{q}}^{\rm{x}}\right)}{D_{\mathrm{q}}^{\rm{x}} \ln \left[\frac{ \ln \left(1 / \varepsilon_{\mathrm{q}}^{\rm{x}}\right)}{\lambda_{\rm{x}} D_{\mathrm{q}}^{\rm{x}}}+1\right]} \text{ packets/s. }
\end{equation}

\subsection{E2E Delay and Packet Loss Probability}
\subsubsection{E2E Path through DA2G Communication}
The E2E delay of DA2G path consists of delay due to backhaul link,  $D_{\mathrm{b}}$, queue delay in the ground BS, $D_{\mathrm{q}}^{\rm{g}}$, and the average transmission delay of the G2A link, $\overline{D}_{\mathrm{t}}^{\rm{ga}}$. Hence, the E2E delay requirement can be satisfied with the following constraint
\begin{equation}
\label{eq.delayG2A}
     D_{\mathrm{b}} + D_{\mathrm{q}}^{\rm{g}} +  \overline{D}_{\mathrm{t}}^{\rm{ga}} \leq D^{\mathrm{th}}.
\end{equation}
By deploying fiber optic backhaul links, we assume that the backhaul delay for remote piloting is around 1 ms\footnote{This value of backhaul delay corresponds to the propagation delay in a path with a length of $300$ km.}.

Correspondingly, the overall packet loss probability is due to the backhaul failure, packet dropping in the ground BS's queue with a probability of $\varepsilon_{\rm{q}}^{\rm{g}}$, and decoding error of the G2A link with a probability of ${\varepsilon}_{\rm{t}}^{\rm{ga}}$. Thus, reliability can be guaranteed if
\begin{equation}
    \label{eq.reliabG2A}
     1-(1-\varepsilon_{\rm{b}})(1-\varepsilon_{\rm{q}}^{\rm{g}})(1-{\varepsilon}_{\rm{t}}^{\rm{ga}})\leq \varepsilon^{\rm{th}}.   
\end{equation}
$\varepsilon_{\rm{b}}$ is the failure probability of backhaul link, which is modeled by a Bernoulli process, and $1-\varepsilon^{\rm{th}}$ is the required reliability.

\subsubsection{E2E Communication Path of JT CoMP}
\begin{figure}[t]
    \centering
            \includegraphics[width=1\columnwidth]{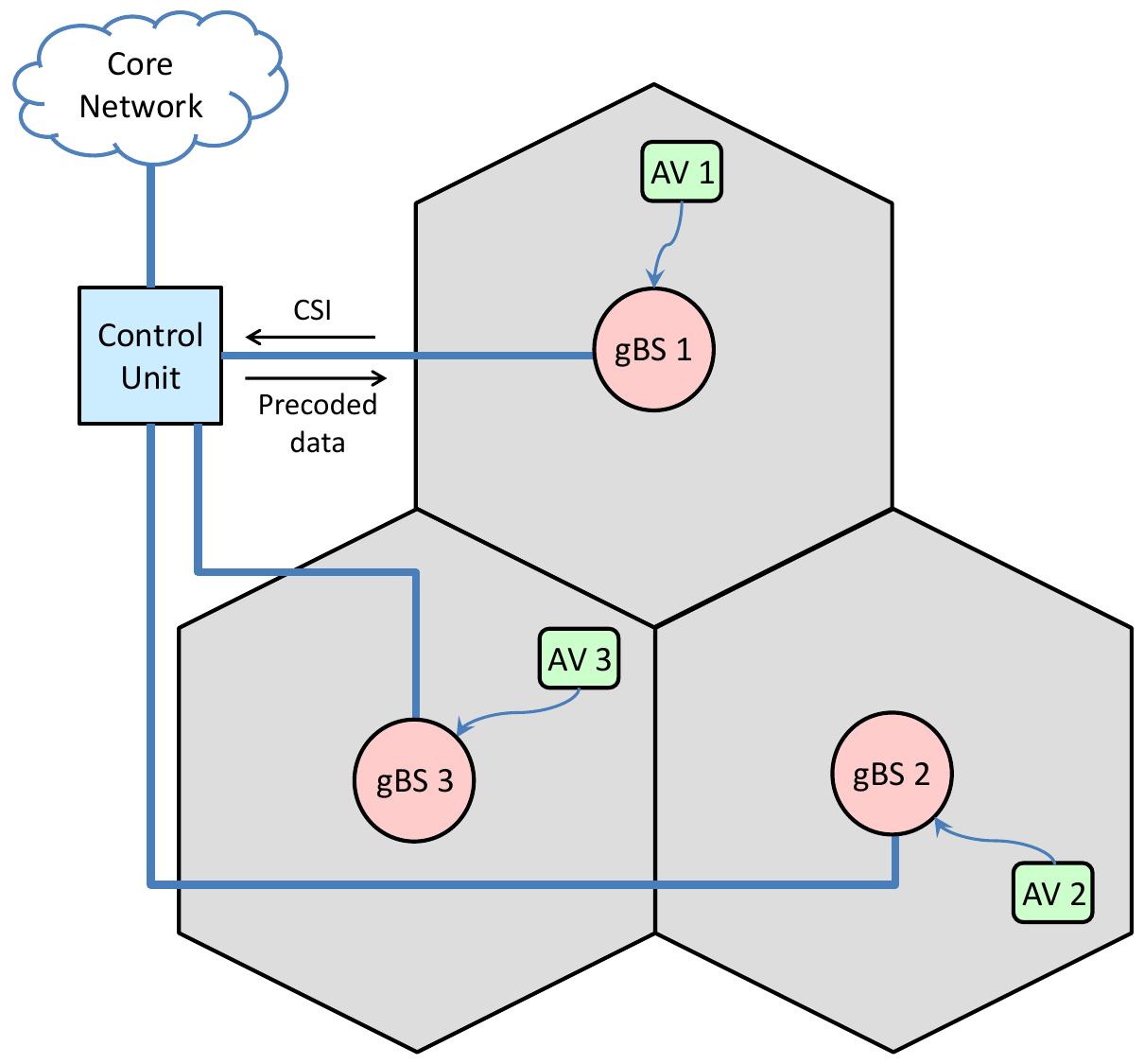}
            \caption{Illustration of centralized CoMP architecture with cluster size of $N=3$.} 
            \label{fig.CoMParc}
\end{figure}

Here, we consider a CoMP cluster, consisting of $N$ ground BSs that are serving $M$ AVs, where $M \leq N$. The E2E delay requirement of JT CoMP with a centralized architecture, introduced in \cite{CoMP_reliability}, is given by 
\begin{equation}
\label{eq.delayJT}
     D_{\mathrm{b}} + D_{\mathrm{c}} + D_{\mathrm{q}}^{\rm{g}} +  \overline{D}_{\mathrm{t}}^{\rm{JT}} \leq D^{\mathrm{th}},
\end{equation}
where $D_{\mathrm{b}}$ as before is the backhaul delay from the core network to the serving ground BSs, and
\begin{equation}
 D_{\mathrm{c}}=\max \limits_{n} \left\{D_{\mathrm{f}}^{\mathrm{g}_n}+D_{\mathrm{b}}^{\rm{C}}+D_{\mathrm{b}}^{\rm{D}}\right\},
\end{equation} 
is the delay due to CoMP, cf. Fig. \ref{fig.CoMParc}, consisting of the delay that AV $m$, $m\in\{1,\cdots,M\}$, feeds back its channel state information (CSI) to its serving BS $n$, $n\in\{1,\cdots,N\}$, i.e., $D_{\mathrm{f}}^{\mathrm{g}_n}$, and the backhaul delay between ground BS $n$ and the control unit (CU) when ground BS $n$ forwards the local CSI to the CU, i.e., $D_{\mathrm{b}}^{\rm{C}}$, and the backhaul delay between CU and ground BS $n$ when the CU distributes precoded data to ground BS $n$, i.e., $D_{\mathrm{b}}^{\rm{D}}$. The feedback delay as in \cite{delay_limited_CoMP} is considered a fixed value of $5$ ms, and we assume the backhaul delay between the ground BS and CU as $D_{\mathrm{b}}^{\rm{C}}=D_{\mathrm{b}}^{\rm{D}}=0.1$ ms\footnote{This value of backhaul delay corresponds to the propagation delay in a distance of $30$ km where BSs are connected with one-hop backhaul \cite{jointUlDl}.}. Moreover, $\overline{D}_{\mathrm{t}}^{\rm{JT}}=\frac{{D_{\rm{t}}^{\rm{ga}}}}{1-{\varepsilon}_{\rm{t}}^{\rm{JT}}}$ is the transmission delay of JT CoMP.

The overall packet loss probability of JT with a CoMP cluster size of $N$ can be calculated as
\begin{equation}
    \label{eq.reliabJT}
     1-(1-\varepsilon_{\rm{b}})(1-\prod_{n=1}^{N}\varepsilon_{\rm{c}}^{\mathrm{g}_n})(1-\prod_{n=1}^{N}\varepsilon_{\rm{q}}^{\mathrm{g}_n})(1-{\varepsilon}_{\rm{t}}^{\rm{JT}})\leq \varepsilon^{\rm{th}},
\end{equation}
where $\varepsilon_{\rm{c}}^{\mathrm{g}_n}$ is the probability that ground BS $n$ fails to cooperate in its CoMP cluster and is given by \cite{CoMP_reliability}
\begin{equation}
\varepsilon_{\rm{c}}^{\mathrm{g}_n}=\varepsilon_{\rm{b}}^{\rm{D}}+(1-\varepsilon_{\rm{b}}^{\rm{D}})\prod_{n=1}^{N}(\varepsilon_{\rm{b}}^{\rm{C}}+(1-\varepsilon_{\rm{b}}^{\rm{C}})\varepsilon_{\rm{f}}^{\mathrm{g}_n}).
\end{equation}
$\varepsilon_{\rm{b}}^{\rm{D}}$ is the failure probability of the backhaul link between the CU and ground BS $n$ when the CU transmits precoded data to ground BS $n$, and $\varepsilon_{\rm{b}}^{\rm{C}}$ is the failure probability of the backhaul link between ground BS $n$ and the CU when ground BS $n$ forwards the local CSI to the CU. $\varepsilon_{\rm{f}}^{\mathrm{g}_n}$
is the link failure probability of
the access link between AV $m$ and ground BS $n$, when the AV feeds back the CSI to ground BS $n$. We suppose that the CSI feedback is error free, i.e., $\varepsilon_{\rm{f}}^{\mathrm{g}_n}\approx0$, so the channel coefficients between all the AVs and their serving ground BSs are perfectly known at the CU. 

Finally, ${\varepsilon}_{\rm{t}}^{\rm{JT}}$ denotes the decoding error probability of JT CoMP and is calculated by  ${\varepsilon}_{\rm{t}}^{\rm{JT}}\approx  Q(f(\gamma^{\rm{JT}},R^{\rm{ga}},D_{\rm{t}}^{\rm{ga}}))$, where $\gamma^{\rm{JT}}$ is the SINR of AV $m$ given by
\begin{equation}
    \label{eq.comp}
    \gamma^{\rm{JT}} = \frac{p_{m}}{P_{\rm{interf}}\sum\limits_{i \in \mathcal{N}_i} p_{i}\left|h_{i}\right|^{2}+B^{\rm{ga}}N_0} \ .
\end{equation} 
$p_{m}$ denotes the symbol power allocated to AV $m$ and based on equal power strategy is derived as \cite{CoMP}
\begin{equation}
    \label{eq.comp_pow}
    p_{m} = \frac{P_{\max}}{\max{[\mathbf{W}\mathbf{W}^*]}_{j,j}} \ .
\end{equation}
$\mathbf{W}$ is the zero-forcing precoding obtained as the pseudo-inverse of the channel matrix, $\mathbf{H} \in \mathbb{C}^{M \times N}$, available at the CU, i.e., $\mathbf{W}=\mathbf{H}^*(\mathbf{H}\mathbf{H}^*)^{-1}$  where $(.)^*$ denotes the conjugate transpose. We assume disjoint CoMP clusters with inter-cluster interference, where $p_{i}$ in \eqref{eq.comp} is the transmit power of interfering BS $i$, with ground BS's power constraint $P_{\max}$. As the worst case of the SINR we assume $p_{i} = P_{\max}$. Since we assume perfect CSI at the CU, the intra-cluster interference due to serving other AVs in the same CoMP cluster is canceled by the zero-forcing precoding.

\subsubsection{E2E Path through A2A Communication}
For the scenario of deploying an AV as a relay to transmit data to the AV of interest, the packet in addition to the DA2G communication path goes across relay AV's queue, with a delay of $D_{\mathrm{q}}^{\rm{a}}$, and A2A link, with an average delay of $\overline{D}_{\mathrm{t}}^{\rm{aa}}$. Hence, the delay components should satisfy
\begin{equation}
\label{eq.delayA2A}
     D_{\mathrm{b}} + D_{\mathrm{q}}^{\rm{g}} + \overline{D}_{\mathrm{t}}^{\rm{ga}} +  D_{\mathrm{q}}^{\rm{a}} + \overline{D}_{\mathrm{t}}^{\rm{aa}} \leq D^{\mathrm{th}}.
\end{equation}

Correspondingly, the reliability of the A2A communication path can be ensured if
\begin{equation}
\begin{split}
    \label{eq.reliabA2A}
     1-(1-\varepsilon_{\rm{b}})(1-\varepsilon_{\rm{q}}^{\rm{g}})(1-{\varepsilon}_{\rm{t}}^{\rm{ga}})(1-\varepsilon_{\rm{q}}^{\rm{a}})(1-{\varepsilon}_{\rm{t}}^{\rm{aa}})\leq \varepsilon^{\rm{th}}.    
\end{split}
\end{equation}

If we consider a swarm of parallel coordinated AVs with single-hop transmission to serve the desired AV with joint decoding strategy, the E2E error probability and delay can be calculated by \eqref{eq.error.tot} and \eqref{eq.delay.tot}, respectively. In fact, it helps increase reliability by exploiting path diversity in the A2A link.

\subsubsection{E2E Path through HAP Communication}
For HAP, long distances of G2H and H2A links cause propagation delay in addition to previous delay components. Therefore, the E2E delay requirement of HAP is satisfied if 
\begin{equation}
\label{eq.delayH2A}
     D_{\mathrm{b}} + D_{\mathrm{q}}^{\rm{g}} + \overline{D}_{\mathrm{t}}^{\rm{gh}} + D_{\mathrm{p}}^{\rm{gh}} + D_{\mathrm{q}}^{\rm{h}} + \overline{D}_{\mathrm{t}}^{\rm{ha}} + D_{\mathrm{p}}^{\rm{ha}} \leq D^{\mathrm{th}},
\end{equation}
where $D_{\mathrm{p}}^{\rm{gh}}$ and $D_{\mathrm{p}}^{\rm{ha}}$ are the propagation delay of the G2H link and the H2A link, respectively. $\overline{D}_{\mathrm{t}}^{\rm{ha}}$ denotes the average transmission delay of the H2A link.

The overall packet loss probability of the HAP communication, similar to the A2A communication, can be computed as
\begin{equation}
\begin{split}
    \label{eq.reliabH2A}
     1-(1-\varepsilon_{\rm{b}})(1-\varepsilon_{\rm{q}}^{\rm{g}})(1-{\varepsilon}_{\rm{t}}^{\rm{gh}})(1-\varepsilon_{\rm{q}}^{\rm{h}})(1-{\varepsilon}_{\rm{t}}^{\rm{ha}})\leq \varepsilon^{\rm{th}}.    
\end{split}
\end{equation}

\subsubsection{E2E Path through LEO Satellite Communication}
The E2E delay constraint of LEO satellite path, similar to the HAP communication, is given by 
\begin{equation}
\label{eq.delayS2A}
     D_{\mathrm{b}} + D_{\mathrm{q}}^{\rm{g}} + \overline{D}_{\mathrm{t}}^{\rm{gs}} + D_{\mathrm{p}}^{\rm{gs}} + D_{\mathrm{q}}^{\rm{s}} + \overline{D}_{\mathrm{t}}^{\rm{sa}} + D_{\mathrm{p}}^{\rm{sa}} \leq D^{\mathrm{th}}.
\end{equation}
where $D_{\mathrm{p}}^{\rm{gs}}$ and $D_{\mathrm{p}}^{\rm{sa}}$ are the propagation delay of the G2S and S2A links, respectively. $\overline{D}_{\mathrm{t}}^{\rm{sa}}$ denotes the average transmission delay of the S2A link.

Due to movement of LEO satellite, in addition to the aforementioned factors, the reliability depends on the availability of LEO satellite links and can be guaranteed if
\begin{equation}
\begin{split}
    \label{eq.reliabS2A}
     1-&(1-\varepsilon_{\rm{b}})(1-\varepsilon_{\rm{q}}^{\rm{g}})(1-\varepsilon_{\rm{l}}^{\rm{gs}})\\
     &(1-{\varepsilon}_{\rm{t}}^{\rm{gs}})(1-\varepsilon_{\rm{q}}^{\rm{s}})(1-\varepsilon_{\rm{l}}^{\rm{sa}})(1-{\varepsilon}_{\rm{t}}^{\rm{sa}})\leq \varepsilon^{\rm{th}}.    
\end{split}
\end{equation}
$\varepsilon_{\rm{l}}^{\rm{xy}}$, $\rm{xy} \in \{\rm{gs},\rm{sa}\}$ is the unavailability probability of LEO satellite X2Y link, which is defined as $1-P_{\mathrm{vis}}^{\rm{xy}}$. Here, we approximate the link availability probability with visibility probability which is given by \cite{LEO}
\begin{equation}
P_{\mathrm{vis}}^{\rm{xy}}=1-\left(1-\frac{{d_{\max }^{\rm{xy}}}^{2}-\hslash_{\mathrm{s}}^{2}}{4 R_{\mathrm{e}}\left(R_{\mathrm{e}}+\hslash_{\mathrm{s}}\right)}\right)^{n_{\mathrm{s}}},
\end{equation}
where $d_{\max }^{\rm{xy}}$ 
is the maximum distance between nodes $\rm{x}$ and $\rm{y}$ at the minimum elevation angle $\vartheta_{\min }$. Moreover, $R_{\mathrm{e}}$ is the Earth radius, $\hslash_{\mathrm{s}}$ 
and $n_{\mathrm{s}}$ are altitude and the number of LEO satellites, respectively.

\section{Best Multi-Connectivity Path Selection}\label{sec.opt}
This section discusses the selection of an MC path that maximizes the E2E spectral efficiency (SE). As redundant connections increase spectrum usage, we focus on SE, the ratio between the effective E2E data rate and the total bandwidth allocated to the desired AV, to minimize the number of required links.
\subsection{Spectrum Efficiency in a Multi-Hop Multi-Connectivity Scenario}
By adopting ARQ scheme, the achieved data rate of the X2Y link can be expressed as
\begin{equation}
    \label{eq.rate.achieved}
    \hat{R}^{\rm{xy}}=\frac{b}{\overline{D}_{\rm{t}}^{\rm{xy}}}\ \text{bits/s},
\end{equation}
where $b$ is the number of bits, and $\overline{D}_{\rm{t}}^{\rm{xy}}$ which is calculated by (24) indicates the average transmission delay of the X2Y link.
In a multi-hop path, the E2E data rate is reflected by the bottleneck link, i.e., the link with the minimum SINR \cite{SE_MH}. Let  $\mathcal{F}_i$ be the set of links in path $i$, hence $\min \left\{ {\hat{R}^{\rm{xy}},\forall {\rm{xy}} \in \mathcal{F}_i  } \right\}$ is the bottleneck rate. On the other hand, redundant connections through MC can lead to an increase in the effective data rate by considering the multi-hop path with the maximum E2E data rate, so as $\max {\left\{ \min \left\{ {\hat{R}^{\rm{xy}},\forall {\rm{xy}} \in \mathcal{F}_i  } \right\},\forall i \in \mathcal{G}_j \right\}}$ where $\mathcal{G}_j$ is the set of paths constituting the MC path $j$ \cite{SE_MC}. Consequently, the SE of MC path $j$ is calculated as
\begin{equation}
    \label{eq.SE}
    {\mathit{SE}_j}=\frac{\max {\left\{ \min \left\{ {\hat{R}^{\rm{xy}},\forall {\rm{xy}} \in \mathcal{F}_i  } \right\},\forall i \in \mathcal{G}_j \right\}}}{\sum_{\forall i \in \mathcal{G}_j} \sum_{\forall {\rm{xy}} \in \mathcal{F}_i}B^{\rm{xy}}}\ \text{bps/Hz}.
\end{equation}
\subsection{Optimization Problem}
To select the best MC path with the minimum required connectivity links that ensure the safe remote piloting operation of BVLoS at a certain level, we formulate the SE maximization problem under the constraints of E2E reliability, E2E delay, and network availability as
\begin{subequations}
\label{eq.opt}
\begin{align}
\mathop {\max }\limits_{j \in \mathcal{H} }& {\;\;\sum_{j=1}^{|\mathcal{H}|} \alpha_j\mathit{SE}_j} \label{eq.a}\\
{\rm{ s.t.}}\quad&{\mathcal{E}_{\rm{E2E}}^j}\leq \varepsilon^{\rm{th}},~\forall j \in \mathcal{H},\label{eq.b}\\
&{\mathcal{D}_{\rm{E2E}}^j}\leq D^{\mathrm{th}},~\forall j \in \mathcal{H},\label{eq.c}\\
&P_{\rm{A}}^j \geq P^{\mathrm{th}},~\forall j \in \mathcal{H},\label{eq.d}\\
& \alpha_j \in \{0,1\}, ~\forall j \in \mathcal{H}, \label{eq.e}\\
& \sum_{j=1}^{|\mathcal{H}|} \alpha_j = 1, ~\forall j \in \mathcal{H} .\label{eq.f}
\end{align}
\end{subequations}
where $\mathcal{H}$ is the set of MC paths, i.e. all the possible combinations of different paths, and $\alpha_j$ is the indicator variable that shows whether the MC path $j$ is selected or not.  Constraints \eqref{eq.b} and \eqref{eq.c} ensure that E2E error probability and E2E delay of MC path $j$ are not greater than the error and delay thresholds, respectively. Constraint \eqref{eq.d} states that the network availability of MC path $j$ is greater than or equal to the required network availability threshold  $P^{\mathrm{th}}$. Constraint \eqref{eq.e} indicates that $\alpha_j$ is a binary variable to show if the MC path $j$ is selected. The last constraint, \eqref{eq.f}, ensures that only one MC path in $\mathcal{H}$ is selected. The optimum path can be found by the brute-force method on the set of MC paths through Monte-Carlo simulation. It should be noted that $\mathit{SE}_j$ in \eqref{eq.a}, is the average value over the number of channel realizations.


\section{Numerical Results}\label{sec.res}

In this section, we evaluate the performance of different E2E connectivity paths comprising multiple RATs and investigate how MC can ensure the stringent requirements of remote piloting the eVTOLs. To this end, we consider an urban scenario with macro cells for the ground network. The system parameters are listed in Table \ref{tab.setup}. The resource blocks (RBs) assigned to each AV consist of 4 consecutive RBs. The subcarrier spacing is $0.2$ MHz. Therefore, the allocated bandwidth of the X2Y link, $B^{\rm{xy}}$, $\rm{xy} \in \{\rm{ga},\rm{aa},\rm{ha},\rm{sa}\}$, to transmit a packet is $0.8$ MHz, which does not exceed the coherence bandwidth $1.2$ MHz \cite{crossLayer_edge}. The dedicated bandwidth of the G2H / G2S link, $B^{\rm{gh}}$ / $B^{\rm{gs}}$, is assumed to be fixed as $1$ MHz. The queueing delay requirement is considered as $D_{\rm{q},\text{max}}=0.7$ ms and $\varepsilon_{\rm{q}}^{\rm{x}}=10^{-6}$ for $\rm{x} \in \{\rm{a},\rm{g},\rm{h},\rm{s}\}$. The average packet arrival rate of AV, ground BS, HAP, and satellite is assumed as $\lambda_{\rm{a}}=100$ packets/s, $\lambda_{\rm{g}}=1000$ packets/s, and $\lambda_{\rm{h}}=\lambda_{\rm{s}}=10000$ packets/s, respectively. So based on \eqref{eq.effBW}, the effective bandwidth of the arrival process to satisfy the queueing delay requirement is determined as $E_{\mathrm{BW}}^{\rm{a}}\approx3700$ packets/s, $E_{\mathrm{BW}}^{\rm{g}}\approx6500$ packets/s, and $E_{\mathrm{BW}}^{\rm{h}}=E_{\mathrm{BW}}^{\rm{s}}\approx18000$ packets/s.
We consider the data rate of all the links, $R^{\rm{xy}}$, as $500$ kbps. In addition, probability of interference, $P_{\rm{interf}}$, and CoMP cluster size, $N$, are assumed $0.05$ and $3$, respectively. In our simulations, the system parameters in most cases are as specified above or listed in Table \ref{tab.setup}, unless otherwise stated.

\begin{table}[t]
\centering
\caption{System Parameters.}
\label{tab.setup}
\vspace{-0.2cm}
\begin{center}
\begin{tabular}{ |l|l| } 
 \hline
 \textbf{System parameter} & \textbf{Value} \\
 \hline
 \hline
 Required reliability, $1-\varepsilon^{\rm{th}}$ & $0.99999$ \\
 \hline
 Delay threshold, $D^{\rm{th}}$ & $20$ ms \\
 \hline
 Packet size, $b$ & $32$ bytes \\ 
 \hline
 Average packet arrival rate of AV, $\lambda_{\rm{a}}$ & $100$ packets/s \cite{jointUlDl}\\ 
 \hline
 Average packet arrival rate of gBS, $\lambda_{\rm{g}}$ & $1000$ packets/s \cite{jointUlDl}\\ 
 \hline
 Average packet arrival rate of HAP, $\lambda_{\rm{h}}$ & $10000$ packets/s \cite{MC-AV}\\ 
 \hline
 Average packet arrival rate of satellite, $\lambda_{\rm{s}}$ & $10000$ packets/s \\ 
 \hline
 Queueing delay bound, $D_{\max}^{\rm{q}}$ & $0.7$ ms \\ 
 \hline
 Queueing delay violation probability, $\varepsilon_{\rm{q}}^{\rm{x}}$ & $10^{-6}$ \\ 
 \hline
 Backhaul failure probability, $\varepsilon_{\rm{b}}$ & $10^{-6}$ \cite{MC-AV}\\
 \hline
 Carrier frequency of all links in S-band, $f_{\rm{c}}$ & $2$ GHz  \\
 \hline
 Carrier frequency of satellite links in Ka-band, $f_{\rm{c}}$ & $30$ GHz  \\
 \hline
 \hline
 AV Tx power & $23$ dBm \cite{3GPP36.777}\\
  \hline
 gBS/HAP Tx power & $46$ dBm \cite{3GPP36.777}\\
  \hline  
 LEO Tx power & $50$ dBm \cite{LEO_small}\\
  \hline 
 AV Tx/Rx antenna gain, $g_{\rm{a}}$ & $0$ dBi \cite{3GPP36.777}\\
  \hline
 Maximum gain of gBS antenna element, $g_{\rm{e}}^{\max}$ & $8$ dBi \cite{U2U_comm}\\
  \hline
 Maximum gain of HAP Tx/Rx antenna, $g_{\rm{h}}^{\max}$ & $32$ dBi \cite{MC-AV}\\
  \hline
 Maximum gain of LEO Tx/Rx antenna, $g_{\rm{s}}^{\max}$ & $38$ dBi \\
  \hline  
 AV Rx noise figure & $9$ dB \cite{3GPP36.777}\\
  \hline
 HAP/LEO Rx noise figure & $5$ dB \cite{3GPP36.777}\\
  \hline  
 Number of gBS antenna elements, $N_{\rm{e}}$ & $8$ \cite{U2U_comm}\\
  \hline
 Downtilt angle, $\phi_{\rm{t}}$ & $102^{\circ}$ \cite{U2U_comm}\\
  \hline
 Inter-site distance (ISD) & $500$ m \cite{3GPP36.777}\\
  \hline
 Height of gBS, $\hslash_{\rm{g}}$ & $25$ m \cite{3GPP36.777}\\
 \hline
 \hline
 Altitude of AV, $\hslash_{\rm{a}}$ & $300$ m \cite{FACOM}\\
  \hline
 Altitude of HAP, $\hslash_{\rm{h}}$ & $20$ km \cite{FACOM}\\
  \hline  
 Altitude of LEO satellite, $\hslash_{\rm{s}}$ & $1110$ km \cite{spaceX} \\
  \hline
 Number of LEO satellites, $n_{\rm{s}}$ & $4425$ \cite{spaceX} \\
  \hline  
 Minimum elevation angle, $\vartheta_{\min }$ & $15^{\circ}$ \cite{3GPP38.811}\\
 \hline
 Rice factor of G2A link, $K_{\rm{ga}}$ & $5\sim12$ dB \cite{MC-AV}\\
  \hline
 Rice factor of A2A link, $K_{\rm{aa}}$ & $12$ dB \cite{MC-AV}\\
  \hline
 Rice factor of G2H link, $K_{\rm{gh}}$ & $5\sim15$ dB \cite{MC-AV}\\
  \hline
 Rice factor of H2A link, $K_{\rm{ha}}$ & $12\sim15$ dB \cite{MC-AV}\\
  \hline
 Rice factor of G2S link, $K_{\rm{gs}}$ & $5\sim15$ dB (S-band)\\ & $10\sim30$ dB (Ka-band) \\
  \hline
 Rice factor of S2A link, $K_{\rm{sa}}$ & $12\sim15$ dB (S-band)\\ & $20\sim30$ dB (Ka-band) \\
  \hline  
 Noise spectral density, $N_0$ & $-174$ dBm/Hz \\
  \hline
 LoS (NLoS) shadow fading standard deviation  & $4$ ($6$) dB \cite{3GPP36.777}\\
  \hline
\end{tabular}
\end{center}
\end{table}

We consider a hexagonal grid for the cellular terrestrial network consisting of $3$ tiers, i.e., $37$ cells in total. $10$ AVs are located randomly with uniform distribution at a fixed altitude over the considered cells. We employ a swarm of at most $3$ coordinated AVs, and $6$ of AVs are interfering with the AV of interest. The location of the desired AV's serving BS and the HAP / LEO satellite projection on the ground is assumed at the origin. The horizontal distance of the HAP (LEO satellite) and its ground station is set as $5$ ($300$) km. Altitude and number of LEO satellites in Table \ref{tab.setup} are assumed based on Starlink constellation. In \cite{Rice_factor}, the Rician $K$-factor was found to increase exponentially with elevation angle between two nodes. Here for simplicity, we assume that the Rician factor of each link increases linearly with the elevation angle. The elevation angles are considered from $0^{\circ}$ to $90^{\circ}$ with a $10^{\circ}$ step, and the Rice factor is assumed to be constant in each interval.
The experiments are provided to assess the reliability and network availability of different E2E paths and their parallel combinations for remote piloting of eVTOLs and investigate how we can achieve high E2E reliability and low E2E latency by MC along with adjusting system parameters such as data rate, bandwidth, CoMP cluster size, and interference level. 

\subsection{Performance Analysis of Single E2E Paths}
\begin{figure}[t]
    \centering
            \includegraphics[width=1\columnwidth]{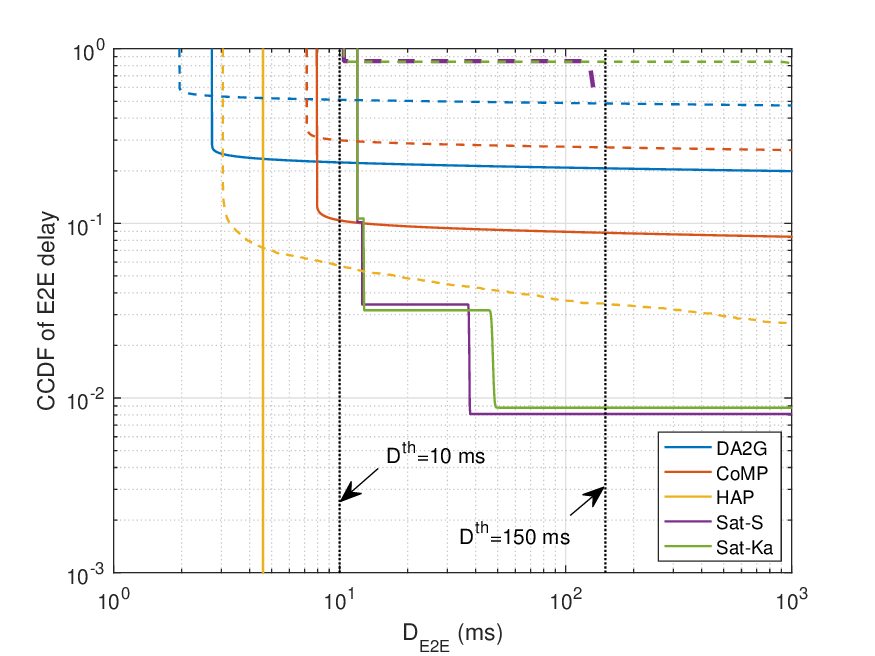}
            \caption{Comparison of CCDF of E2E delay in different RATs. The solid lines and the dash lines represent the CCDF with data rate of $250$ kbps and $1$ Mbps, respectively.}
            \label{fig.CCDf}
\end{figure}

First, we analyse the E2E delay of different RATs. In Fig. \ref{fig.CCDf}, we draw the complementary cumulative distribution function (CCDF) of different paths' E2E delay, meaning the probability that E2E delay is greater than abscissa. 
It is observed that the least E2E delay is provided by DA2G and then by HAP communication. While JT CoMP results in more E2E delay due to coordination among ground BSs. Finally, LEO satellite, in both S-band and Ka-band, incur the most E2E latency owing to the large propagation delay. The solid lines that indicate the CCDF of E2E delay with the lowest required data rate of remote piloting, i.e., $250$ kbps, reveal that DA2G, HAP, CoMP, and LEO Satellite with high probability result in E2E latency of $\simnot3$ ms, $\simnot5$ ms, $\simnot8$ ms, and $\simnot10$ ms, respectively. For the highest required data rate, i.e., $1$ Mbps, since the transmission delay decreases, the dash lines indicate improvement of the E2E delay performance of different paths. It is observed that E2E delay of LEO satellite always exceeds the minimum threshold which is $10$ ms.
We note that the horizontal asymptote of latency CCDF is equal to the packet drop probability, according to the delay-based reliability in \eqref{eq.delayReliabil}. It is clearly seen in Fig. \ref{fig.CCDf} that there is a trade-off between the latency and reliability. It is observed that with increasing data rate from $250$ kbps to $1$ Mbps, the packet drop probability of DA2G (JT CoMP) increases from $\simnot0.2$ ($\simnot0.1$) to $\simnot0.5$ ($\simnot0.3$). For HAP (LEO satellite), it grows dramatically from $\simnot0$ ($\simnot0.01$) to $\simnot0.03$ ($\simnot0.8$).

\begin{figure}[!htb]
\centering
    \begin{subfigure}
        \centering
        \includegraphics[width=1\columnwidth]{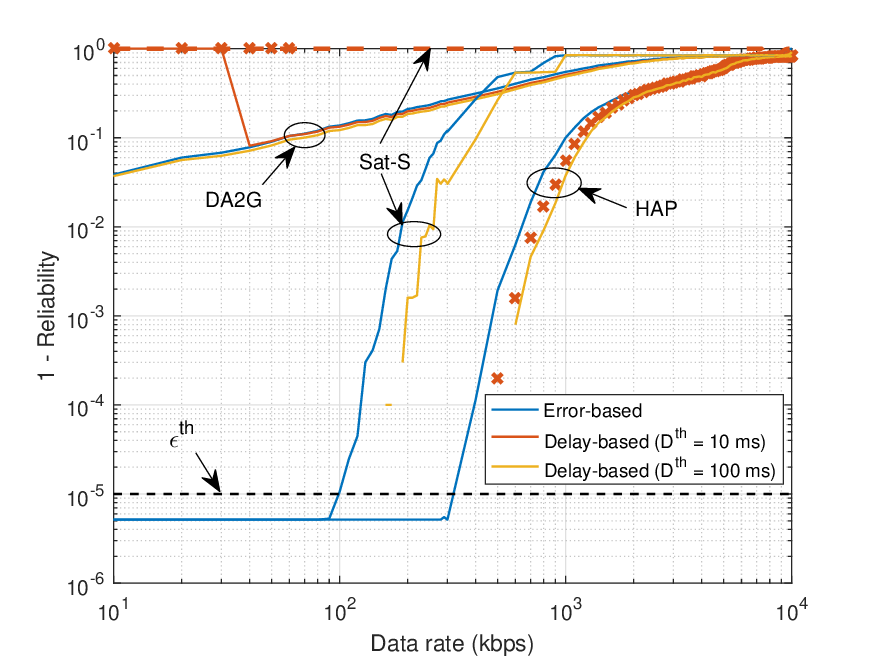}\\ \footnotesize{(a)}
    \end{subfigure}
    \vfill
    \begin{subfigure}
        \centering
        \includegraphics[width=1\columnwidth]{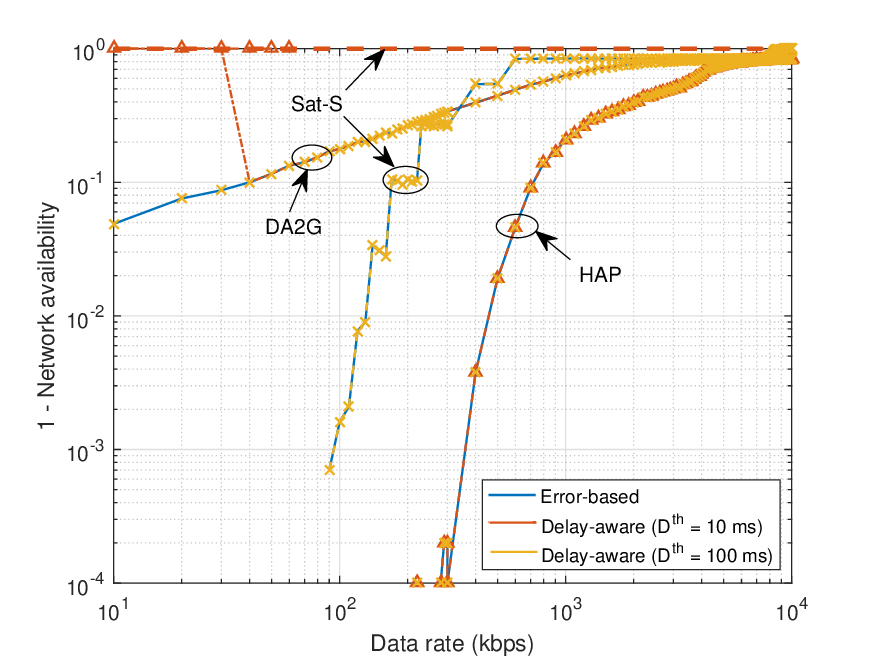}\\ \footnotesize{(b)}
    \end{subfigure}
    \vfill
    \caption{Comparison of (a) reliability and (b) network availability based on two definitions in different RATs.}
    \label{fig.definition}
\end{figure}

In Fig. \ref{fig.definition}(a), we show the relationship between two reliability definitions in Section \ref{sub.sysKPI}, i.e., error-based reliability in \eqref{eq.errorReliabil} and delay-based reliability in \eqref{eq.delayReliabil}. It is observed that these two definitions are almost similar in high data rates that decoding error is the dominant factor of packet dropping. Moreover, delay-based reliability depends on $D^{\rm{th}}$, such that the lower the delay threshold, the higher the packet drop probability. On the opposite, in low data rates that transmission delay increases and so E2E delay exceeds $D^{\rm{th}}$, while decoding error probability is low, the gap between the two definitions is huge. About DA2G and HAP communications, when $D^{\rm{th}}=10$ ms, for data rates lower than $\simnot40$ kbps and $\simnot70$ kbps, respectively, which E2E delay is more than threshold, the performance gap grows. For LEO satellite, based on the previous results in Fig. \ref{fig.CCDf}, since E2E delay always exceeds $10$ ms, the packet drop probability with $D^{\rm{th}}=10$ ms is $1$. It is observed that for certain data rate intervals in HAP/satellite communication, there is no value for delay-based reliability. Because in none of $10$ million realizations of the experiment, E2E delay did not exceed the desired threshold.
Fig. \ref{fig.definition}(b) indicates the relationship of the conventional network availability without delay constraint, i.e. error-based availability defined in \eqref{eq.errorAvailabil}, and with delay constraint, named delay-aware availability as in \eqref{eq.availability}. It is obvious that they are equivalent unless E2E delay violates $D^{\rm{th}}$. Hence, with strict delay threshold of $10$ ms, as previous graph, in low data rates a gap arises between the two definitions.

Furthermore, from Fig. \ref{fig.definition}(a), it is realized that in the range of desired data rates from $0.25\sim 1$ Mbps, the reliability of DA2G and LEO satellite communication is not higher than $\simnot0.8$ and $\simnot0.99$, respectively, which are in accordance with the results in Fig. \ref{fig.CCDf}. At the same time, based on Fig. \ref{fig.definition}(b), their network availability is less than $\simnot0.8$ and $\simnot0.9$, respectively, which is not acceptable for the C2 application. On the other hand, HAP communication is the most reliable and available path that can satisfy the target reliability of $0.99999$ with network availability as high as $\simnot0.9999$ just up to $\simnot300$ kbps. The empirical results verify that single path can not satisfy the stringent requirements, individually. Thus, in the following subsections, we evaluate the key performance metrics, i.e., reliability and network availability of multi-path connectivity by equations \eqref{eq.errorReliabil} and \eqref{eq.availability} with respect to some system parameters.

\subsection{Impact of Data Rate on MC Performance}
\begin{figure}[t]
\centering
    \begin{subfigure}
        \centering
        \includegraphics[width=1\columnwidth]{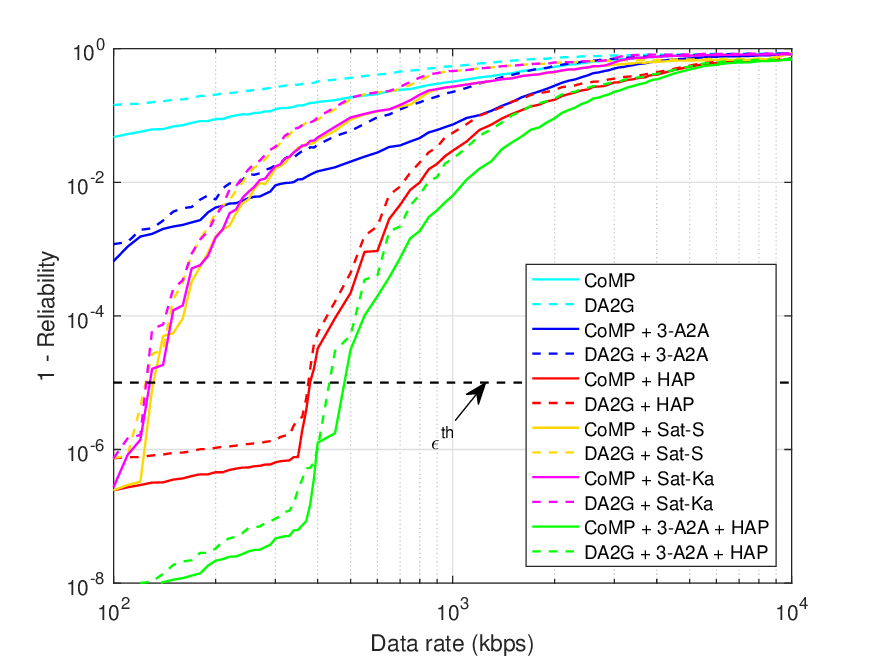}\\ \footnotesize{(a)}
    \end{subfigure}
    \vfill
    \begin{subfigure}
        \centering
        \includegraphics[width=1\columnwidth]{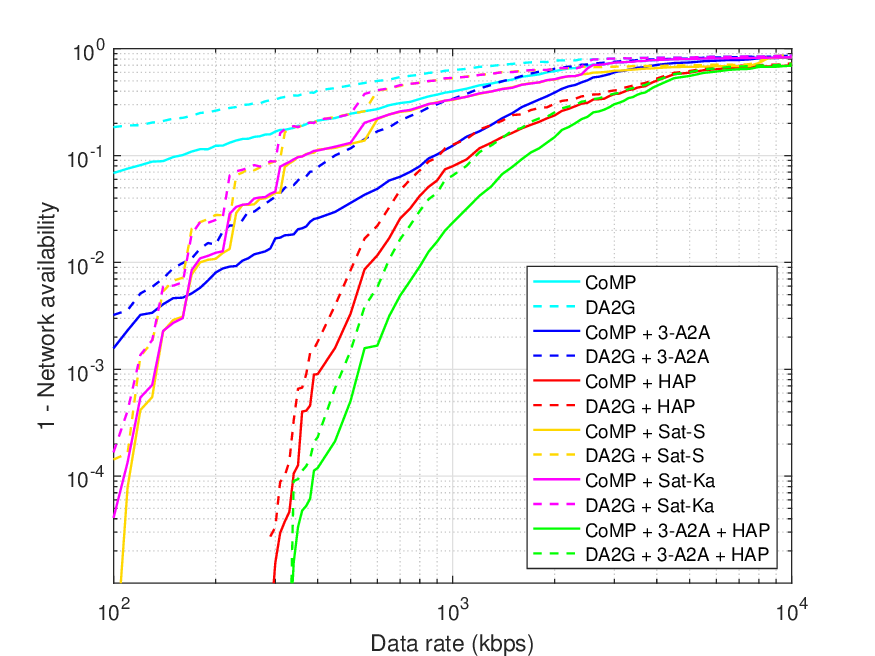}\\ \footnotesize{(b)}
    \end{subfigure}
    \vfill
    \caption{(a) Reliability and (b) network availability performance of multi-path connectivity vs. data rate.}
    \label{fig.MCrate}
\end{figure}

Fig. \ref{fig.MCrate} shows the overall error probability and network unavailability of different multi-path connectivity with respect to the data rate when the AVs' allocated bandwidth, $B^{\rm{xy}}$, $\rm{xy} \in \{\rm{ga},\rm{aa},\rm{ha},\rm{sa}\}$, is $0.8$ MHz. CoMP cluster size and probability of interference are set as $3$ and $0.05$, respectively. Fig. \ref{fig.MCrate} depicts the performance gain of multiple communication paths connectivity with DA2G / JT CoMP as a master connectivity. It is observed that for the minimum required data rate of $250$ kbps, the reliability of ``DA2G + 3-A2A" and ``DA2G + Sat-S/Ka" schemes is $\simnot0.99$, and their network availability is $\simnot0.97$ and $\simnot0.93$, respectively, which shows improvement compared to the single RAT transmission. Furthermore, ``DA2G + HAP" and ``DA2G + 3-A2A + HAP" schemes improve the target reliability of $0.99999$ with network availability of $\simnot0.999$ up to $\simnot400$ kbps and $\simnot500$ kbps data rates, respectively.
Additionally, it is shown that JT CoMP improves the reliability and network availability compared with DA2G communication because of combating the inter-cell interference by cooperation among ground BSs. The results show the cooperation of $3$ adjacent ground BSs. For further improvements we can increase the CoMP cluster size, as its effect is investigated in the next subsection.

\subsection{Impact of CoMP Cluster Size}
\begin{figure}[t]
\centering
    \begin{subfigure}
        \centering
        \includegraphics[width=1\columnwidth]{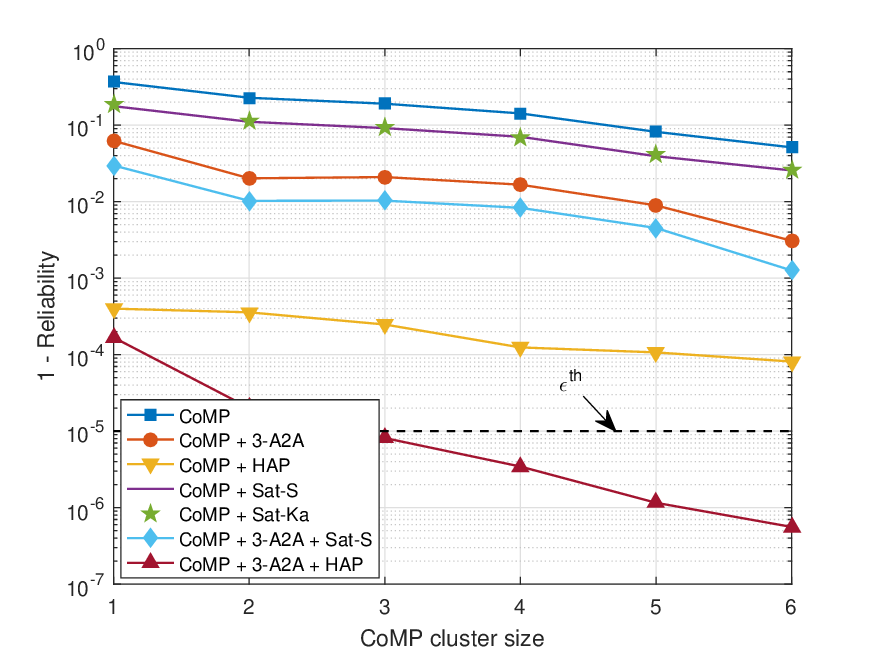}\\ \footnotesize{(a)}
    \end{subfigure}
    \vfill
    \begin{subfigure}
        \centering
        \includegraphics[width=1\columnwidth]{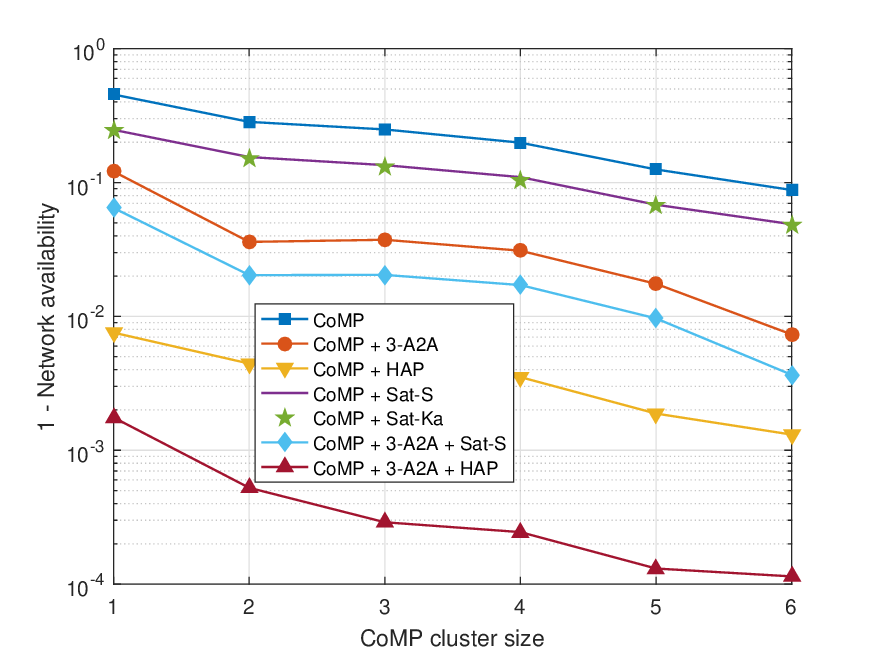}\\ \footnotesize{(b)}
    \end{subfigure}
    \vfill
    \caption{(a) Reliability and (b) network availability performance vs. CoMP cluster size.}
    \label{fig.comp}
\end{figure}

In Fig. \ref{fig.comp}, we investigate how the CoMP cluster size affects the reliability and network availability, when data rate and AV's allocated bandwidth are $500$ kbps and $0.8$ MHz, respectively, and $P_{\rm{interf}}=0.05$. As shown in Fig. \ref{fig.comp}, the reliability and availability can be improved by increasing CoMP cluster size. In this figure, CoMP cluster size of $1$ is equivalent to DA2G communication. The performance gap between the cluster size of $1$ and $2$, i.e., adopting DA2G or JT CoMP, is notable, especially when A2A links via JT CoMP are considered as the auxiliary communication path, such as ``CoMP + 3-A2A", ``CoMP + 3-A2A + Sat-S/Ka", and ``CoMP + 3-A2A + HAP" schemes. Thus, utilizing JT CoMP along with A2A links and increasing CoMP cluster size can be a promising approach to achieve the target reliability and network availability. As it is observed, ``CoMP + 3-A2A + HAP" scheme with cluster size of at least $3$ can achieve the required reliability in the evaluated scenario.

\subsection{Effect of the Bandwidth Allocation}
\begin{figure}[t]
\centering
    \begin{subfigure}
        \centering
        \includegraphics[width=1\columnwidth]{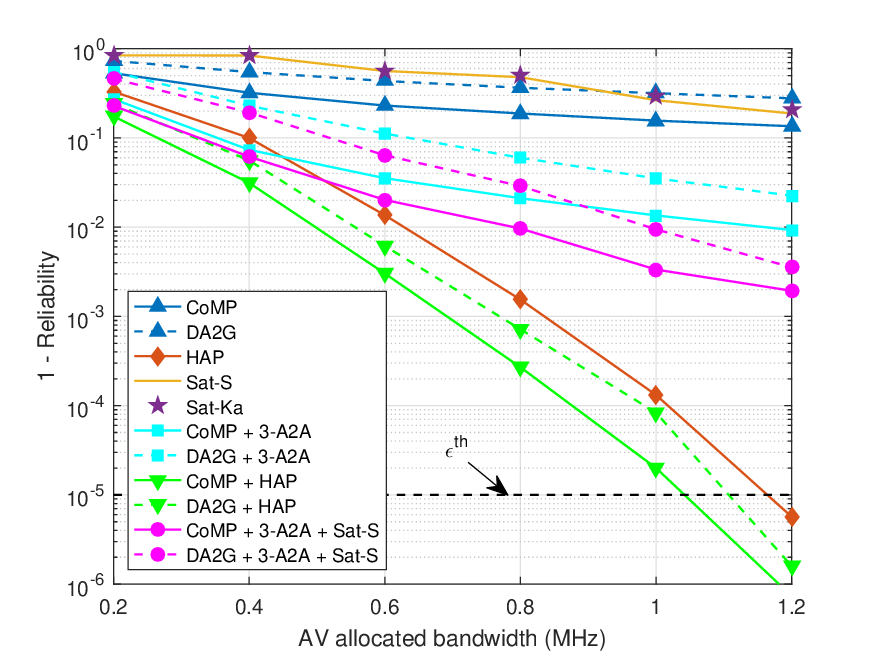}\\ \footnotesize{(a)}
    \end{subfigure}
    \vfill
    \begin{subfigure}
        \centering
        \includegraphics[width=1\columnwidth]{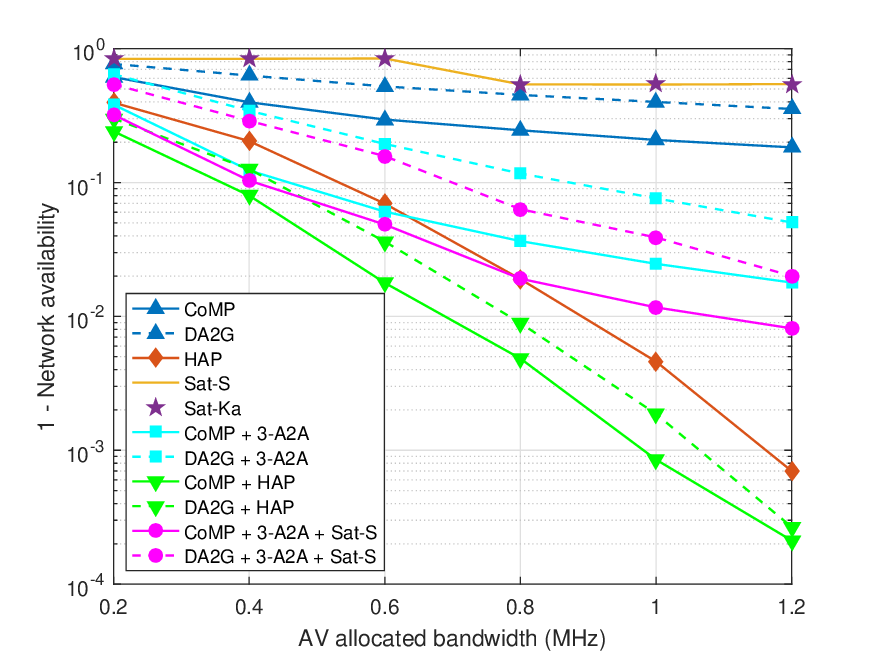}\\ \footnotesize{(b)}
    \end{subfigure}
    \vfill
    \caption{(a) Reliability and (b) network availability performance vs. AV's allocated bandwidth.}
    \label{fig.BW}
\end{figure}

The relation between the performance metrics, i.e., the reliability and network availability, and the AV's allocated bandwidth is illustrated in Fig. \ref{fig.BW}. The bandwidth of one RB is $0.2$ MHz, and the total bandwidth allocated to each AV does not exceed the coherence bandwidth of $1.2$ MHz. So, at most $6$ consecutive RBs can be assigned to each AV. Unlike single paths of DA2G, CoMP, and satellite communication which seems not to achieve significant improvement in reliability and availability with respect to the allocated bandwidth, multi-path connectivity and especially HAP benefit significantly from this aspect. It is observed that HAP communication can individually achieve the target reliability of $0.99999$, and availability of $\simnot0.999$ with allocating $6$ RBs, while both of these values are less than $\simnot0.9$ with assigning $1$ RB.

\subsection{Effect of Interference}
\begin{figure}[t]
\centering
    \begin{subfigure}
        \centering
        \includegraphics[width=1\columnwidth]{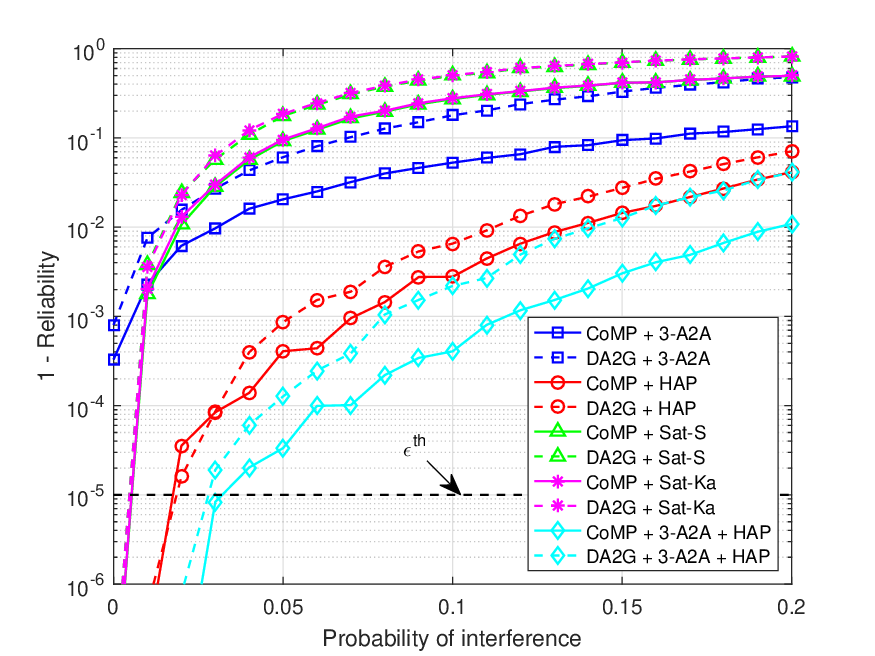}\\ \footnotesize{(a)}
    \end{subfigure}
    \vfill
    \begin{subfigure}
        \centering
        \includegraphics[width=1\columnwidth]{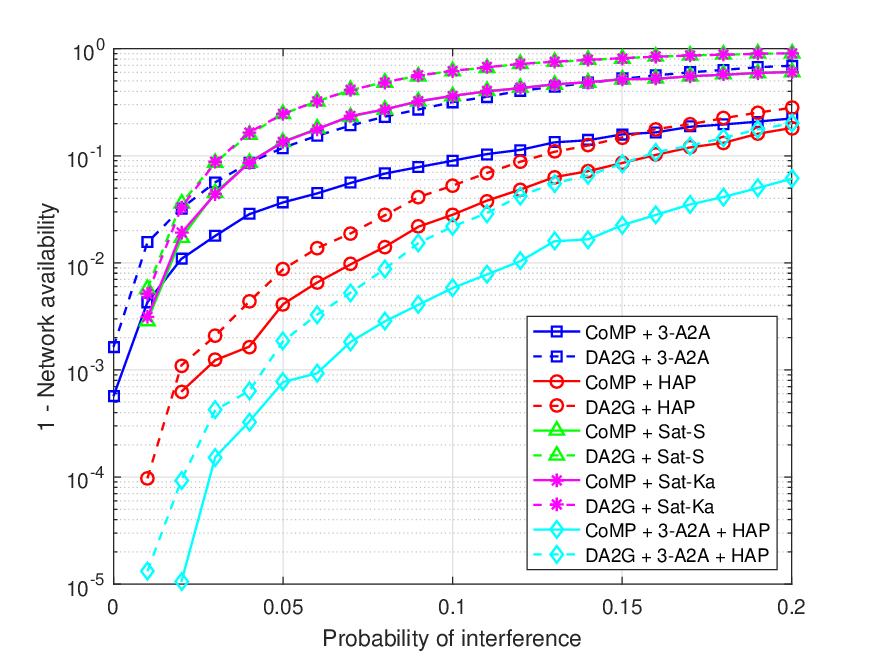}\\ \footnotesize{(b)}
    \end{subfigure}
    \vfill
    \caption{(a) Reliability and (b) network availability performance vs. probability of interference.}
    \label{fig.interf}
\end{figure}

In Fig. \ref{fig.interf}, we examine the effect of interference on the performance of different links and MC schemes, when data rate, bandwidth, and CoMP cluster size are $500$ kbps, $0.8$ MHz, and $3$, respectively. For each RAT, we assume particular frequency band with full frequency reuse such that each X2Y link, $\rm{xy} \in \{\rm{ga},\rm{aa},\rm{ha},\rm{sa}\}$, incurs interference with probability of $P_{\rm{interf}}$ from all the corresponding links. The results in Fig. \ref{fig.interf}(a) show that DA2G and A2A links are highly interference limited due to LoS paths even in very low probabilities such as $0.001$. Additionally, satellite S/Ka-band's performance becomes rapidly saturated with interference. As an example, by increasing the probability of interference from $0.001$ to $0.2$ the reliability of ``DA2G + Sat-S/Ka" and ``CoMP + Sat-S/Ka" schemes degrades from higher than $6$-nines ($1-10^{-6}$) to $\simnot0.2$ and $\simnot0.5$, respectively. Finally, HAP's performance diminishes gradually from higher than $6$-nines to $\simnot0.93$ and $\simnot0.96$, in ``DA2G + HAP" and ``CoMP + HAP" schemes, respectively, by increasing the interference probability from $0.001$ to $0.2$. Moreover, it is observed that with probability of interference greater than $0.03$, none of the considered multiple paths can provide the target reliability of $0.99999$ and network availability higher than $0.9999$ in the evaluated scenario.

\subsection{Best MC Path Selection}
\begin{figure}[t]
    \centering
            \includegraphics[width=1\columnwidth]{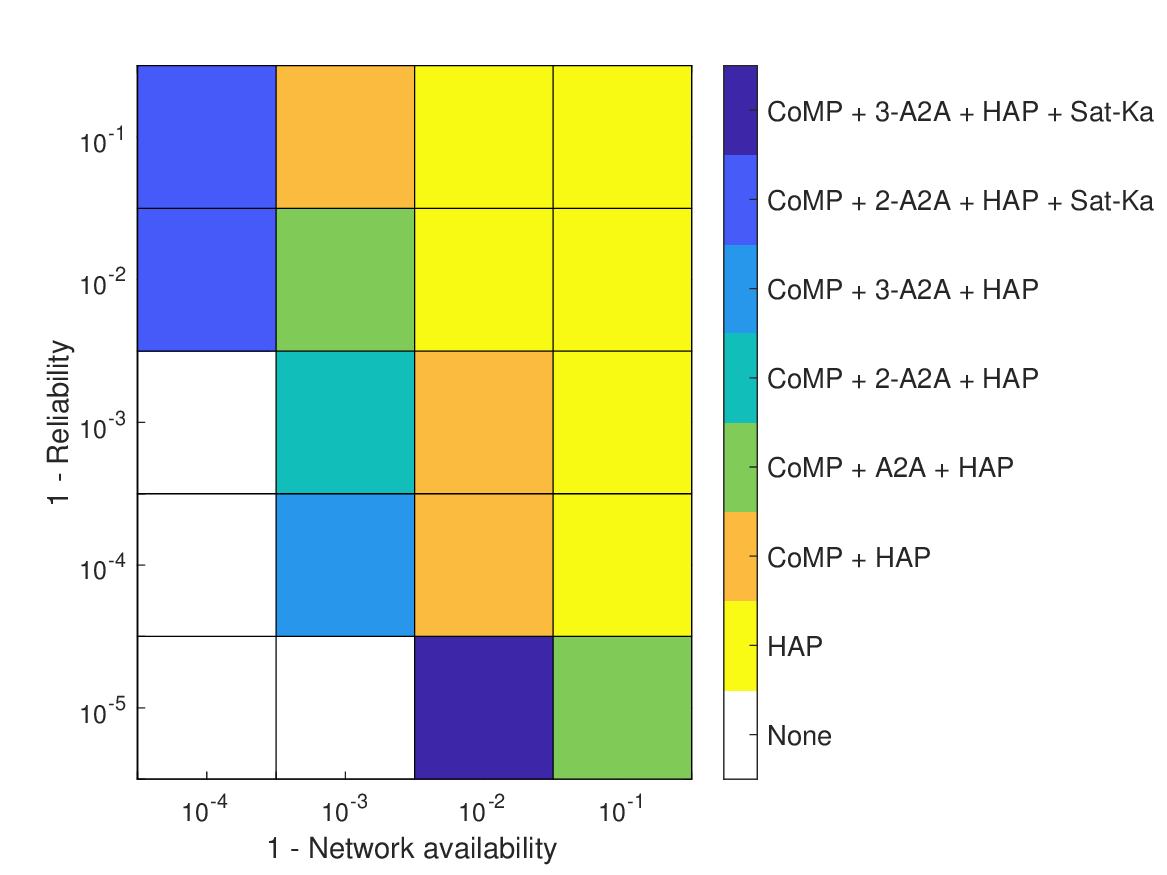}
            \caption{Best MC path with the minimum required links for different reliability and network availability demands.}
            \label{fig.opt}
\end{figure}

In Fig. \ref{fig.opt}, we determine the optimal MC path with the minimum required links that fulfill the E2E delay of $20$ ms under diverse E2E reliability and network availability requirements. The amount of data rate and the allocated bandwidth of different links are considered as $500$ kbps and $0.8$ MHz, respectively. The probability of interference and CoMP cluster size are $0.05$ and $3$, respectively. From this graph, it is observed that HAP communication solely is enough to fulfill the E2E reliability of $0.9999$ with the target network availability of $0.9$. Also, it can guarantee the reliability of $0.99$ with the target network availability of $0.99$. For higher reliability and/or network availability requirements, the optimum MC scheme demands more number of multiple paths. It is observed that a combination of all the RATs, i.e., ``CoMP + 3-A2A + HAP + Sat-Ka", is able to ensure the target reliability of $0.99999$ under the network availability of $0.99$. Furthermore, it is observed that there are some specific cases that there is no MC path in the experiment to guarantee the service requirements. Such high reliability and network availability demand other investigations of design parameters such as bandwidth, CoMP cluster size, and effective interference mitigation techniques.

\section{Conclusion}\label{sec.con}
In this paper, we have studied the beyond visual line-of-sight (BVLoS) of remote piloting an aerial vehicle (AV) in finite blocklength (FBL) regime with multi-connectivity (MC) under practical antenna configurations. To this end, we have integrated multi radio access technologies (RATs) including direct air-to-ground (DA2G), air-to-air (A2A), high altitude platform (HAP), and low Earth orbit (LEO) satellite communications. A major challenge of DA2G communication is the management of severe line-of-sight (LoS) interference. Coordinated multi-point (CoMP) in joint transmission (JT) mode is a well known technique to overcome inter-cell interference, since base stations (BSs) cooperatively process signals. Hence, we exploit JT CoMP to improve the performance gain. Overall packet loss probability and end-to-end (E2E) latency are characterized as functions of the system parameters such as required data rate, AV's allocated bandwidth, CoMP cluster size, probability of interference, and backhaul failure probability. We evaluate the reliability, delay, and network availability of multiple communication path connectivity for command and control (C2) link. We have shown that the overall performance of different links under practical antenna settings is highly limited due to the LoS interference. Moreover, we have demonstrated that even with  interference mitigation techniques, such as JT CoMP, MC is a key enabler for safe operation of a special type of AVs, i.e, electric vertical take-off and landing vehicles (eVTOLs). Moreover, we explored different MC options in order to figure out how to adjust system parameters to provide the quality of service requirements of the mission-critical scenario. 
Finally, we solved an optimization problem to select the best MC path under the service requirements constraints. We maximized spectral efficiency (SE) to specify the optimum MC path with the minimum number of required links and alleviate the spectrum usage of the MC scheme.
As future work, we will investigate new approaches to fulfill higher service requirements. Moreover, the effect of mobility of AVs and the blocking probability of wireless channels by clouds and rain will be studied.

\bibliographystyle{IEEEtran}
\bibliography{paper.bib}

\end{document}